\documentclass[a4paper,hyperref,oneside]{cernrep} 
\pdfoutput=1
\usepackage{snow_title}

\usepackage{tocloft}
\usepackage{texnames}
\usepackage[T1]{fontenc}
\usepackage[square,numbers,sort]{natbib}
\usepackage{url}
\usepackage{hyperref}
\usepackage{booktabs}
\usepackage{subfig}
\usepackage{graphicx}
\usepackage{feynmp}
\usepackage{textpos} % for placement of logo 
\usepackage{relsize} 

\usepackage{mathptmx}
\usepackage[scaled=.90]{helvet}
\usepackage{courier}
\usepackage[T1]{fontenc} %font enconding
\usepackage[latin9]{inputenc} %input encoding

\usepackage{enumitem}
\setlist{noitemsep}
\setlist[itemize]{leftmargin=*, topsep=-2pt, partopsep=0pt, itemsep=2pt}
\setlist[enumerate]{leftmargin=*, topsep=0pt, partopsep=0pt, itemsep=0pt}

\renewcommand{\contentsname}{Table of Contents}

\newcommand{\bibliocommand}{%
}

\newtoks{\abstract}{}
\newtoks{\authors}{}
\newtoks{\affils}{}

\newcommand{\ednote}[1]{}

\newcommand{\tabt}[1]{\multicolumn{1}{c}{#1}}
\newcommand{\tabtt}[1]{\multicolumn{2}{c}{#1}}
\newcommand{\tabttt}[1]{\multicolumn{3}{c}{#1}}

\newcommand{\halfwidth}{0.49\textwidth}
\newcommand{\leftfig}{left}
\newcommand{\rightfig}{right}

\newcommand{\Leftfig}{Left}
\newcommand{\Rightfig}{Right}

\newcommand{\clicsid }{CLIC\_SiD\xspace}
\newcommand{\clicild }{CLIC\_ILD\xspace}

\newcommand{\gghadrons}{\mbox{\ensuremath{\upgamma\upgamma \rightarrow \mathrm{hadrons}}}\xspace}
\newcommand{\gghadron}{\mbox{\ensuremath{\upgamma\upgamma \rightarrow \mathrm{hadron}}}\xspace}
\newcommand{\roots }{\ensuremath{\sqrt{s}}\xspace}
\newcommand{\GeV }{\ensuremath{\text{GeV}}\xspace}
\newcommand{\MeV }{\ensuremath{\text{MeV}}\xspace}
\newcommand{\TeV }{\ensuremath{\text{TeV}}\xspace}
\newcommand{\mZ }{\ensuremath{m_{\PZ}}\xspace}
\newcommand{\mH }{\ensuremath{m_{\PH}}\xspace}
\newcommand{\GH }{\ensuremath{\Gamma_{\PH}}\xspace}
\newcommand{\gHZZ}{\ensuremath{g_{\PH\PZ\PZ}}\xspace}
\newcommand{\gHWW}{\ensuremath{g_{\PH\PW\PW}}\xspace}

\newcommand{\gHtt}{\ensuremath{g_{\PH\PQt\PQt}}\xspace}
\newcommand{\gHbb}{\ensuremath{g_{\PH\PQb\PQb}}\xspace}
\newcommand{\gHcc}{\ensuremath{g_{\PH\PQc\PQc}}\xspace}

\newcommand{\gHTauTau}{\ensuremath{g_{\PH\PGt\PGt}}\xspace}
\newcommand{\gHMuMu}{\ensuremath{g_{\PH\PGm\PGm}}\xspace}

\newcommand{\kHZZ}{\ensuremath{\kappa_{\PH\PZ\PZ}}\xspace}
\newcommand{\kHWW}{\ensuremath{\kappa_{\PH\PW\PW}}\xspace}
\newcommand{\kHtt}{\ensuremath{\kappa_{\PH\PQt\PQt}}\xspace}
\newcommand{\kHbb}{\ensuremath{\kappa_{\PH\PQb\PQb}}\xspace}
\newcommand{\kHcc}{\ensuremath{\kappa_{\PH\PQc\PQc}}\xspace}
\newcommand{\kHTauTau}{\ensuremath{\kappa_{\PH\PGt\PGt}}\xspace}
\newcommand{\kHMuMu}{\ensuremath{\kappa_{\PH\PGm\PGm}}\xspace}
\newcommand{\kHGluGlu}{\ensuremath{\kappa_{\PH\Pg\Pg}}\xspace}
\newcommand{\kHGamGam}{\ensuremath{\kappa_{\PH\PGg\PGg}}\xspace}

\newcommand{\rootsprime }{\ensuremath{\sqrt{s^\prime}}\xspace}
\renewcommand{\pT}{\ensuremath{p_\mathrm{T}}\xspace}
\newcommand{\abinv}{\ensuremath{\mathrm{ab}^{-1}}\xspace}
\newcommand{\fbinv}{\ensuremath{\mathrm{fb}^{-1}}\xspace}

\newcommand{\LumiInt}{\ensuremath{\mathcal{L}_{\mathsf{int}}}\xspace}

\newcommand{\epem}{\ensuremath{\Pep\Pem}\xspace}
\newcommand{\mpmm}{\ensuremath{\PGmp\PGmm}\xspace}  %mu+mu-
\newcommand{\ttbar}{\ensuremath{\PQt \PAQt}\xspace}

\newcommand{\qqqqWW}{\ensuremath{\PQq\PAQq\PQq \PAQq\PWp\PWm}\xspace}

\newcommand{\bbbb} {\ensuremath{\PQb\PAQb\PQb\PAQb}\xspace}
\newcommand{\tbtb} {\ensuremath{\PQt\PAQb\PQb\PAQt}\xspace}

\newcommand{\ww}{\ensuremath{\PWp\PWm}\xspace} %WW
\newcommand{\modelone}{\textit{model~I}\xspace}

\newcommand{\modelthree}{\textit{model~III}\xspace}
\newcommand{\susy}[1]{\ensuremath{\widetilde{\mathrm{#1}}}}%             
\newcommand{\slepton}[1]{\ensuremath{\susy{\ell}^{#1}}}
\newcommand{\chargino}[1]{\ensuremath{\PSGc_1^{#1}}}
\newcommand{\neutralino}[1]{\ensuremath{\PSGc_{#1}^0}}
\newcommand{\nene}[2]{\ensuremath{\neutralino{#1}\neutralino{#2}}}  %neutralino_i, neutralino_j
\newcommand{\chch}[2]{\ensuremath{\PSGc_{#1}^{+}\PSGc_{#2}^{-}}}  %chargino_i,chargino_j

\DeclareRobustCommand{\PSGmpR}{\HepSusyParticle{\PGm}{R}{+}\Xspace}

\DeclareRobustCommand{\PSepR}{\HepSusyParticle{\Pe}{R}{+}\Xspace}

\DeclareRobustCommand{\PSHz}{\HepParticle{\PSH}{}{0}\Xspace} % H-zero 1

\DeclareRobustCommand{\PepR}{\HepParticle{\Pe}{R}{+}\Xspace} 
\DeclareRobustCommand{\PemR}{\HepParticle{\Pe}{R}{-}\Xspace}

\DeclareRobustCommand{\PepL}{\HepParticle{\Pe}{L}{+}\Xspace} 
\DeclareRobustCommand{\PemL}{\HepParticle{\Pe}{L}{-}\Xspace}

\DeclareRobustCommand{\PSGtpDo}{\HepSusyParticle{\PGt}{1}{+}\Xspace} % stau
\DeclareRobustCommand{\PSGtDo}{\HepSusyParticle{\PGt}{1}{}\Xspace} % stau

\newcommand{\fbarf}{\HepParticle{f}{}{}\,\HepAntiParticle{f}{}{}\xspace}

\newcommand{\Zprime }{\ensuremath{\PZ^\prime}\xspace}

\newcommand{\micron}{\ensuremath{\upmu\mathrm{m}}\xspace}

\begin{document}

\title{Physics at the CLIC e$^{+}$e$^{-}$ Linear Collider \\ {\relsize{-1} Input to the Snowmass process 2013}}

\authors{The CLIC Detector and Physics Study\\
H.~Abramowicz\affiliated{1}, A.~Abusleme\affiliated{2}, K.~Afanaciev\affiliated{3}, G.~Alexander\affiliated{1}, N.~Alipour Tehrani\affiliated{4}, O.~Alonso\affiliated{5}\textsuperscript{,}\affiliated{6}, K.K.~Andersen\affiliated{7}, S.~Arfaoui\affiliated{4}, C.~Balazs\affiliated{8}\textsuperscript{,}\affiliated{9}, T.~Barklow\affiliated{10}
% ...
}%
\affils{%
\let\thefootnote\relax\footnotetext[1]{\textsuperscript{$\star$} Corresponding Editors. Contact: \href{mailto:clicsnow-editors@cern.ch}{clicsnow-editors@cern.ch}}\affiliation[1]{School of Physics and Astronomy, Faculty of Exact Sciences, Tel Aviv University, Tel Aviv, Israel}, \affiliation[2]{Pontificia Universidad Cat\'{o}lica de Chile, Santiago de Chile, Chile}, \affiliation[3]{National Scientific and Educational Centre of Particle and High Energy Physics, Belarusian State University, Minsk, Belarus}, \affiliation[4]{CERN, Geneva, Switzerland}, \affiliation[5]{Spanish Network for Future Linear Colliders}, \affiliation[6]{Universitat de Barcelona, Barcelona, Spain}, \affiliation[7]{Aarhus University, Aarhus, Denmark}, \affiliation[8]{Australian Collaboration for Accelerator Science (ACAS)}, \affiliation[9]{Monash University, Melbourne, Australia}, \affiliation[10]{SLAC, Stanford, USA},
% ...
}%

%maketitle % title is now in introduction

%% This is a special command, not commonly usable!
\abstract{This paper summarizes the physics potential of the CLIC high-energy \epem linear collider.  It provides input to the Snowmass 2013 process for the energy-frontier working groups on The Higgs Boson (HE1), Precision Study of Electroweak Interactions (HE2), Fully Understanding the Top Quark (HE3), as well as The Path Beyond the Standard Model -- New Particles, Forces, and Dimensions (HE4). It is accompanied by a paper describing the CLIC accelerator study, submitted to the Frontier Capabilities group of the Snowmass process \cite{Dannheim:2013ypa}.}

%\tableofcontents
\makeatletter
\let\thetitle\@title
\makeatother

\begin{textblock*}{0mm}(135mm,-18mm)
  \noindent\includegraphics[width=3cm]{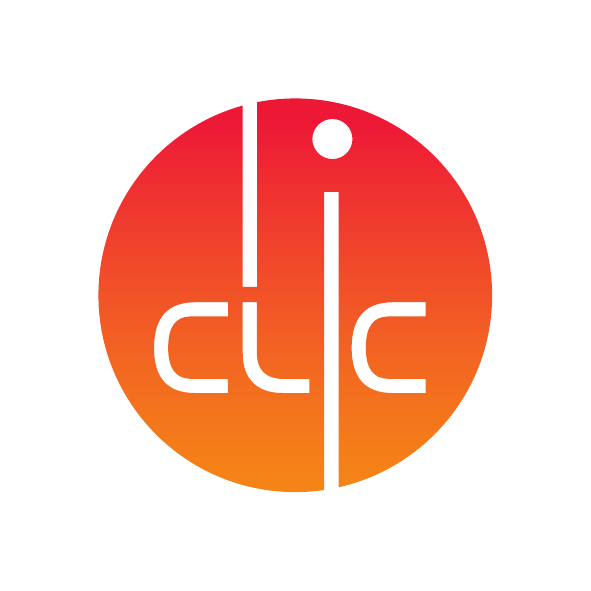}% 
\end{textblock*}%
\vspace{-2cm}
\begin{center}\vspace{-2\parskip}
  \huge \thetitle
\end{center}
\begin{center}
  \today
\end{center}
\noindent
\the\abstract
\author{\the\authors}
{ \affiliations{\the\affils}}%

\tableofcontents

%%% Local Variables: 
%%% mode: latex
%%% TeX-master: "../clicsnowmass"
%%% TeX-PDF-mode: t 
%%% End: 

\section{Introduction}
\label{sec:introduction}

The Compact Linear Collider (CLIC) is a TeV scale high-luminosity linear \epem collider under development.
It is based on a novel two-beam acceleration technique providing acceleration gradients at the
level of 100 MV/m. Recent implementation studies for CLIC have converged towards a staged approach
offering a unique physics program spanning several decades. In this scheme, CLIC would provide
high-luminosity \epem collisions from a few hundred GeV to 3 TeV. The first stage, at or
above the 350 GeV top-pair-production threshold, gives access to precision Higgs physics through the
Higgs-strahlung and WW-fusion production processes, providing absolute values of Higgs couplings to
both fermions and bosons. This stage also addresses precision top physics. The second stage, around
1.4~TeV, opens the energy frontier, allowing for the discovery of New Physics phenomena. This stage
also gives access to additional Higgs properties, such as the top-Yukawa coupling, the Higgs potential
and rare Higgs decay branching ratios. The ultimate CLIC energy of 3 TeV enlarges the CLIC physics
potential even further, covering the complete scope for precision Standard Model physics, direct searches
for pair-production of new particles up to 1.5~TeV mass and optimal sensitivity to New Physics
at much higher mass-scales through precision measurements. A staged implementation of CLIC along the
lines described would open the door to an impressive long-term physics program at the
energy frontier, beyond the LHC program. The machine is therefore considered an important option
for a post-LHC facility at CERN, as emphasized in the recent update of the
European Strategy for Particle Physics \cite{esc16,esu2013}.

Over the last years, the feasibility studies for the CLIC accelerator have systematically and successfully
addressed the main technical challenges of the accelerator project. Similarly, detailed detector and
physics studies confirm the ability to perform high-precision measurements at CLIC.

For more detailed descriptions we refer to the following documents:
\begin{itemize}
%\item CLIC \epem Linear Collider Studies, eds. D. Dannheim et al., submitted to
%  the update process of the European Strategy for Particle Physics, July 2012~\cite{CLIC_strategy_input};
\item A Multi-TeV Linear Collider based on CLIC Technology, CLIC
  Conceptual Design Report, 2012, eds.  M. Aicheler et al.~\cite{CLICCDR_vol1};
\item Physics and Detectors at CLIC, CLIC Conceptual Design Report, eds. L. Linssen et al.~\cite{CLIC_PhysDet_CDR};
\item The CLIC Programme: towards a staged \epem Linear Collider exploring the
  Terascale, CLIC Conceptual Design Report, 2012, eds. P. Lebrun et al.~\cite{CLICCDR_vol3};
\item The Physics Case for an \epem Linear Collider, eds. J. Brau et al.,
  submitted to the update process of the European Strategy for Particle Physics,
  July 2012~\cite{LC_Phys_Case_Strategy_Input}.
\end{itemize}  
\vspace{\parskip}
The CLIC Conceptual Design Report (CDR) is supported by more than 1300
signatories\footnote{\url{https://edms.cern.ch/document/1183227/}} from the
world-wide particle physics community.

%%% Local Variables: 
%%% mode: latex
%%% TeX-master: "../clicsnowmass"
%%% TeX-PDF-mode: t 
%%% End: 

\subsection{CLIC Accelerator Parameters and Options for a Staged Implementation}

The CLIC accelerator design is based on a novel two-beam acceleration scheme. It uses a high-intensity drive beam to generate RF power at 12~GHz. The RF is used to accelerate the main particle beam that runs in parallel to the drive beam. CLIC uses normal-conducting accelerator structures, operated at room temperature. The initial drive beams and main beams are generated in central complexes and are then injected at the end of the two-beam linac arms. The feasibility of the CLIC accelerator has been demonstrated through prototyping, simulations and large-scale tests, as described in the conceptual design report~\cite{CLICCDR_vol1}.  In particular, the two-beam acceleration at gradients exceeding 100~MV/m has been demonstrated in the CLIC test facility CTF3. High luminosities are achieved by very small beam emittances, which are generated in the injector complex and maintained during transport to the interaction point. 

The CLIC accelerator can be built in energy stages, re-using the existing equipment for each new stage. At each energy stage the center-of-mass energy can be tuned to lower values within a range of approximately a factor three with limited loss in luminosity performance. The ultimate choice of the CLIC energy stages will be driven by the physics aims, where further input from LHC data, in particular 14~TeV data, is expected. The recent LHC Higgs discovery makes an initial energy stage around 350~GeV to 375~GeV very attractive, but final choices will depend on further LHC findings. In the first stage around 350~GeV and second stage around 1.4~TeV a single drive-beam generation complex feeds both linacs, while in the third stage at 3~TeV each linac is fed by a separate complex. The accelerator parameters are based on detailed accelerator studies, described in the CDR~\cite{CLICCDR_vol1}, and are used for the physics studies presented in this paper. The assumed integrated luminosities of 0.5~\abinv at 350~GeV, 1.5~\abinv at 1.4~TeV, and 2.0~\abinv at 3~TeV correspond each to four or five years of operation of a fully commissioned machine running 200 days per year with an effective up-time of 50\% (\autoref{tab:accParameter}). The CLIC design foresees 80\% electron polarization, while space is reserved in the layout for a positron polarization option. 

Motivated by the discovery of the 125~GeV Higgs boson, studies for a klystron-based initial stage at 375~GeV are currently being carried out. This option could provide a faster implementation, while still allowing for the re-use of equipment at the higher energy stages.

\begin{table}[btp]
  \centering
  \caption{Center-of-mass energy and assumed integrated luminosity for the different
    CLIC machine stages. The integrated luminosities correspond each to four or five
    years of operation of a fully commissioned machine running 200 days per year
    with an effective up-time of 50\%.}\label{tab:accParameter}
%% Fraction for 350 and 1400 are calculated from the lumi.ee.out files in the clic/machine/beam-beam afs space
  \begin{tabular}{l c c cccc c}
    \toprule
    Parameter                                 & Symbol   & Unit   & Stage 1 & Stage 2 & Stage 3 \\\midrule
    Center-of-mass energy                     & $\roots$ & GeV    & 350     & 1400    & 3000    \\
    Integrated luminosity                     & \LumiInt & \abinv & 0.5     & 1.5     & 2.0     \\
    \bottomrule
  \end{tabular}
\end{table}

\subsection{CLIC Detectors}
\label{sec:clic-detectors}

The detector concepts used for the CLIC physics studies are based on the
SiD~\cite{Aihara:2009ad,ilctdrvol4:2013} and ILD~\cite{ildloi:2009,ilctdrvol4:2013} detector concepts for the
International Linear Collider. They were adapted for the CLIC 3~TeV
accelerator stage, which constitutes the most challenging environment for the
detectors. In a staged scenario, most sub-detectors will serve at all center-of-mass energies, while \eg the inner tracking and vertex detectors would profit from a version with a smaller inner radius at the lower energies. 

\begin{figure}[tb]
  \centering
  \includegraphics[scale=1,clip]{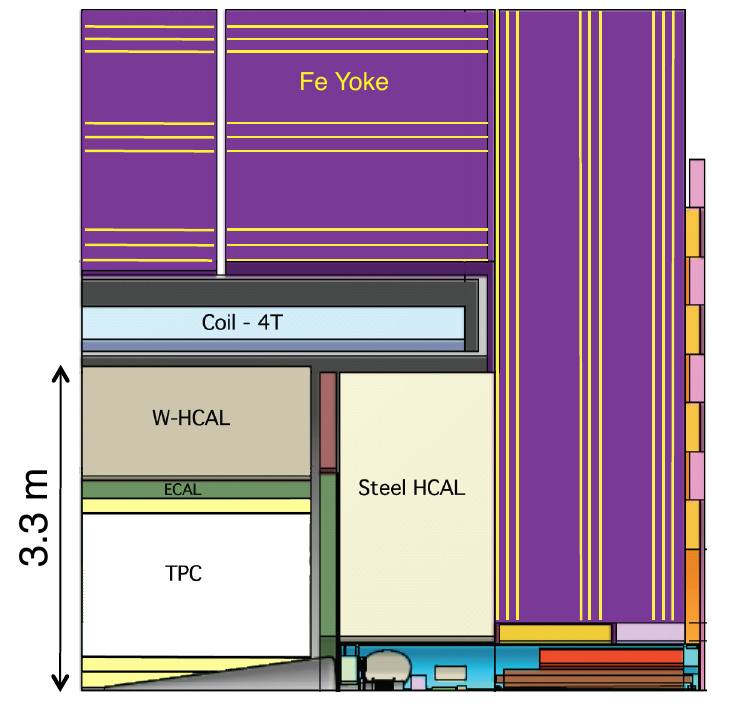}
  \includegraphics[scale=1,clip]{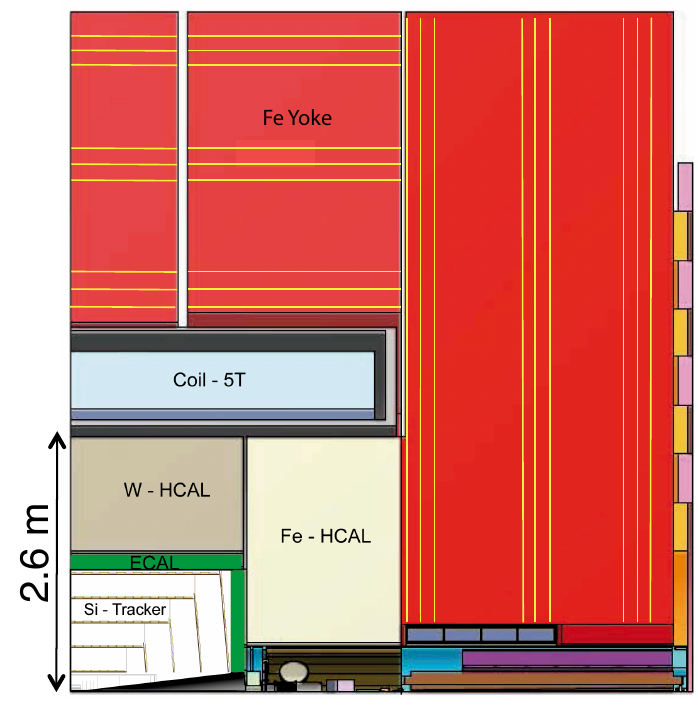}
  % \subfloat[]{\includegraphics[scale=1,clip]{introduction/CLIC_SiD_xz_scale}}\\
  % \subfloat[]{\includegraphics[scale=1,clip]{introduction/clicild}}
  \caption{Longitudinal cross section of the top right quadrant of \clicild (left) and \clicsid (right).}
  \label{fig:detectors}
\end{figure}

\subsubsection{Detector Performance Requirements and Detector Concepts}
\label{sec:detect-requ}

The performance requirements for the CLIC detectors are given by the physics
goals. They are: 
\begin{itemize}
 \item Jet energy resolution of $\sigma_E/E \lesssim 3.5\%$ for jet energies from 100~GeV to 1~TeV  ($\lesssim 5\%$ at 50~GeV);
 \item Track momentum resolution of $\sigma_{\pT}/\pT^2 \lesssim
   2\cdot 10^{-5}~\textrm{GeV}^{-1}$;
 \item Impact parameter resolution  of $\sigma^{2}_{d_{0}}=(5~\micron)^{2}+(15~\micron)^{2}/p^{2}\sin^{3}{\theta}$;
 \item Lepton identification efficiency better than $95\%$ over the full range of energies; 
 \item Detector coverage for electrons down to very low angles. 
\end{itemize}
\vspace{0.2cm}
The jet energy resolution is required to distinguish \PW, \PZ, or \PH bosons, for example, to study the decay chains of charginos and neutralinos.
The momentum resolution is driven, for example, by the muon momentum measurement in HZ recoil events or for measuring the Higgs decay to muons. It is also motivated by, for example, slepton mass measurements.    
Efficient vertex reconstruction and flavor tagging is needed, for example for the measurement of Higgs branching ratios to beauty or charm. 
% It calls for a high-resolution low-mass vertex detector, with small pixels.
% The multiple scattering term of 15~\micron implies that the material budget is
% smaller than 0.2\%~\xo per layer or less than the equivalent of
% 200~\micron of silicon for the active material, readout, support and cooling combined. 
The full coverage is important to suppress Standard Model backgrounds in various physics analyses. In particular, many physics background processes involve electrons in the forward region.

Vertical cuts  through the top-right quadrant of \clicild (left) and \clicsid (right) are shown in \autoref{fig:detectors}. The jet energy resolution requirement is the main
driver of the detector concept designs. As a result both detector
concepts are based on fine-grained calorimeters and optimized for particle-flow
analysis techniques. In the particle-flow approach
all visible particles are reconstructed by combining the information from precise tracking with highly granular calorimetry. The technique achieves an optimal jet energy resolution through the separation of individual particles within the jet~\cite{thomson:pandora,Marshall2013153}. The detectors comprise strong central solenoid magnets, with a field of 5~T in \clicsid and 4~T in \clicild. The tracking system of \clicsid is fully based on silicon pixel and strip detectors, while the tracker of \clicild combines silicon pixels with silicon strips and a large Time Projection Chamber.

CLIC beams will arrive at the detector in bunch trains, occurring every 20 ms. Each bunch train generates 312 bunch crossings at 0.5~ns time separation (3~TeV values). This time structure allows for a trigger-less
readout of the detectors after each bunch train. It also allows for a power-pulsing scheme of the on-detector electronics, thereby significantly reducing the power dissipation and the tracker mass. On average less than one physics event per bunch train is expected. However, the high CLIC energies and small intense beams lead to significant beamstrahlung, resulting in high rates of incoherent electron--positron pairs and \gghadron events. The energy loss through beamstrahlung also generates a tail to the luminosity spectrum that extends well below the nominal center-of-mass energy. 

\begin{figure}[tb]
\centering
\includegraphics[width=\halfwidth]{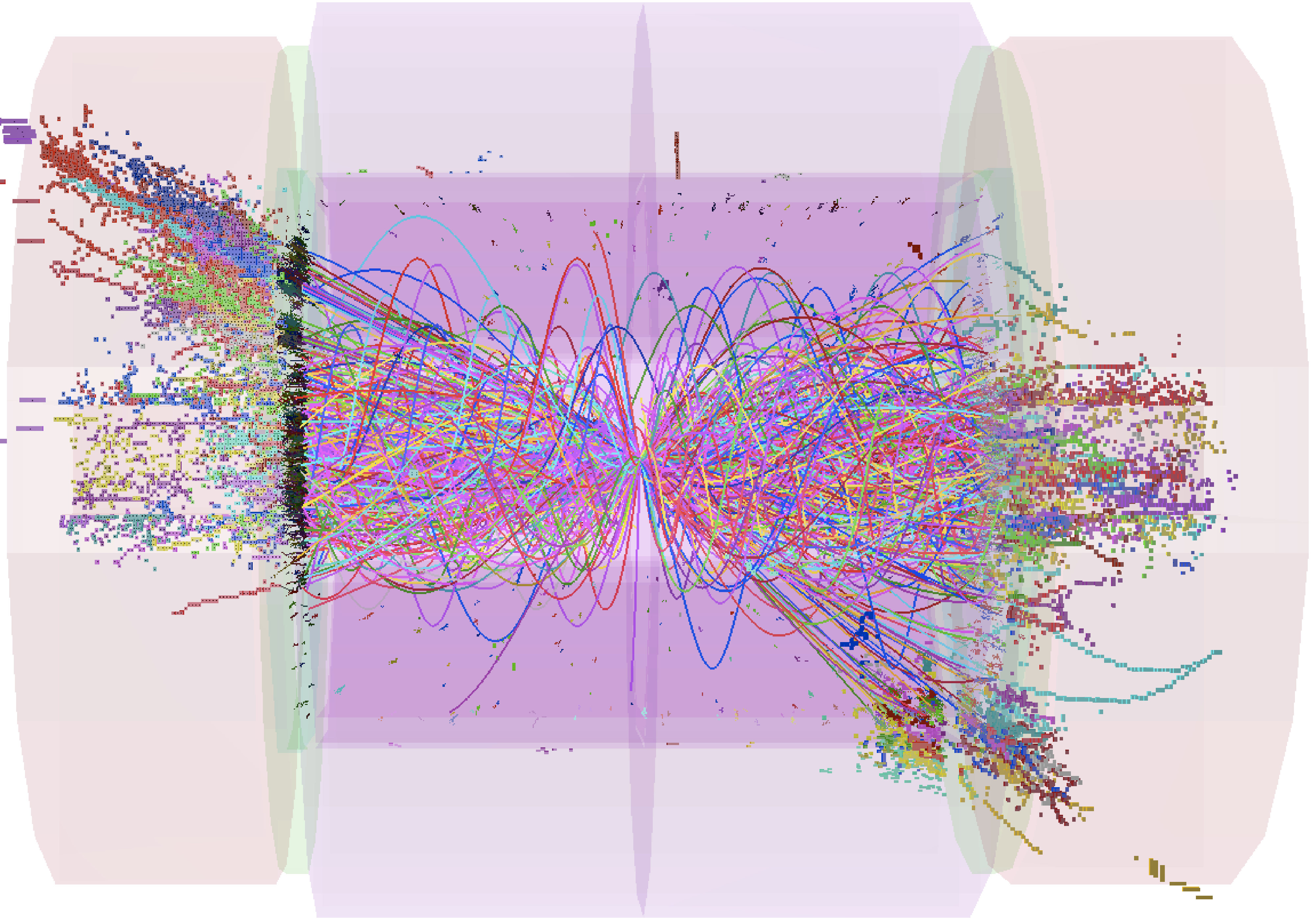}
\includegraphics[width=\halfwidth]{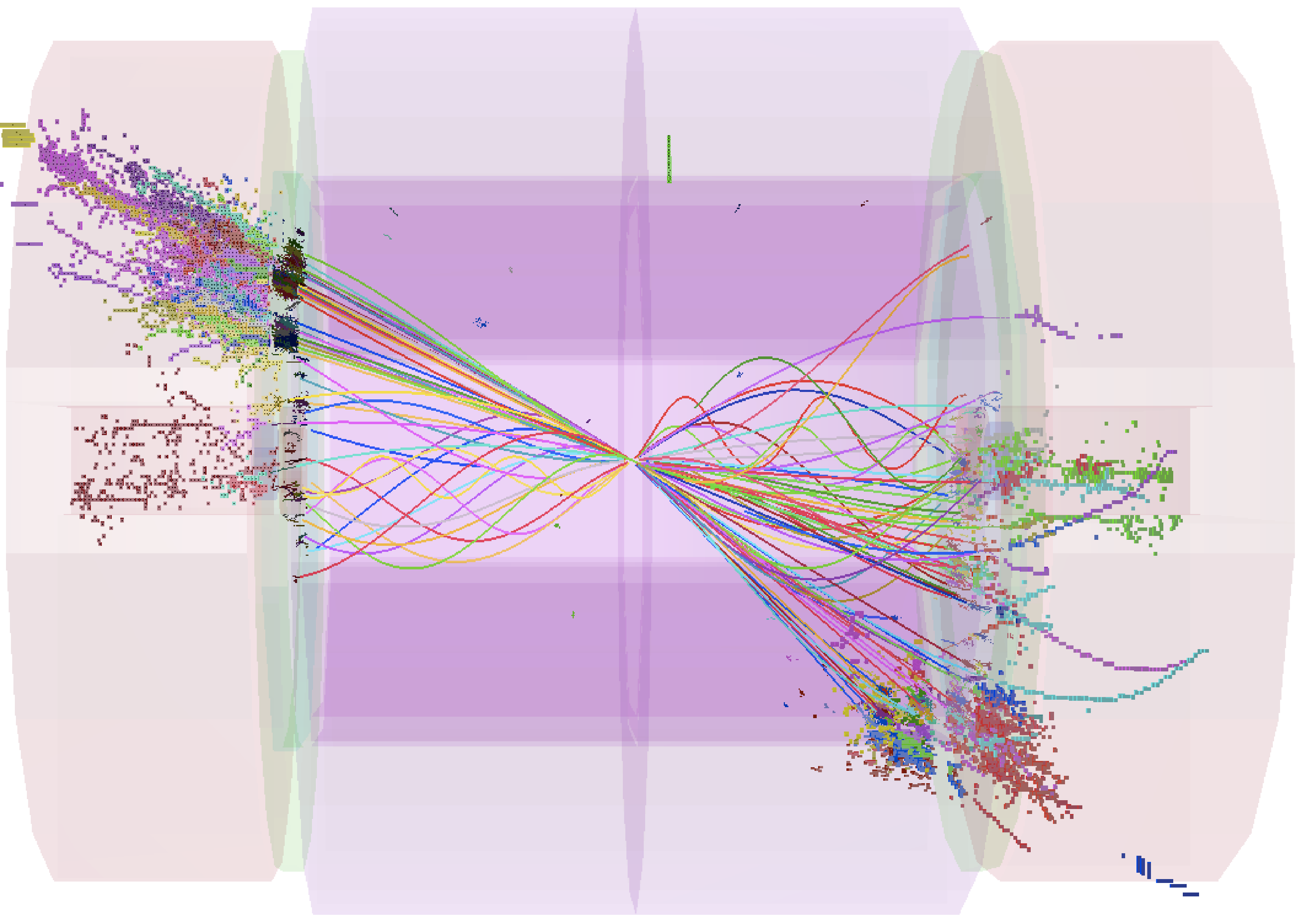}
\caption{Left: Reconstructed particles in a simulated $\epem \to \ttbar $ event
  at 3~TeV in the \clicild detector concept with background from \gghadrons
  overlaid in a broad reconstruction window around the physics event. Right: The effect of applying tight timing cuts on the reconstructed
  particles. \ednote{Replace with SiD like event displays.}}
\label{fig:ttEventDisplay}
\end{figure}

% \subsubsection{Rejection of Beam-induced Backgrounds}
Even at 3~TeV, the impact of beam-induced backgrounds can be reduced by making use of the high
spatial and temporal granularity provided by the detectors. The method relies on precise hit timing (10~ns time-stamping for all silicon tracking elements and 1~ns hit time resolution
for all calorimeter hits) combined with offline event reconstruction, including the background particles, with particle flow analysis. With a tight set of cuts applied
to the reconstructed low-\pT particles, the average background level can
be reduced from approximately 20~TeV per bunch train to about 100~GeV per
reconstructed physics event. This background rejection, which is exemplified in
\autoref{fig:ttEventDisplay}, is achieved without significantly impacting the
physics performance. The remaining background particles can be further rejected
by applying hadron-collider type jet clustering algorithms, which treat the very
forward particles similar to the underlying event in hadronic collisions.

The CLIC CDR studies have demonstrated that the requirements for high-precision physics measurements under CLIC experimental conditions can be fulfilled. This is illustrated in more detail in the following sections. 

%-------------------------------------------------------------------

%%% Local Variables: 
%%% mode: latex
%%% TeX-master: "../clicsnowmass"
%%% TeX-PDF-mode: t 
%%% End: 

\subsection{CLIC Physics Optimization with Energy Staging}
\label{sec:physics-staging}

The main asset and aim of the CLIC acceleration technology is to provide scope
for exploring New Physics at multi-TeV \epem center-of-mass energies and with
very high luminosities. Until now it is the only technology option on the market
that can reach such high energies and for which the feasibility has been
demonstrated in large-scale test facilities and at the level of the design
parameters required for the complete facility, as detailed in the conceptual
design report~\cite{CLICCDR_vol1}. The CLIC technique requires large investments
in its injector complex, while offering a relatively modest marginal cost per GeV for the
acceleration linac~\cite{CLICCDR_vol3}. CLIC is therefore principally justified
as an exploration machine for New Physics at the high-energy frontier. The full
physics potential of CLIC, however, also includes known SM physics, with highly
relevant topics like precision Higgs and top physics to be studied from
$\roots$ of a few hundred GeV onwards. New Physics could already show up at the
scale of a few hundred GeV, as \epem collisions may give access to states, \eg
electroweak particles, that could remain undetected in proton-proton
collisions at the LHC\@. At increasing $\roots$, additional Higgs production and
decay channels become accessible, while the window for New Physics opens up more
and more. 

Cross sections for many of the interesting phenomena are low, at the fb
level, therefore high integrated luminosities are essential all along the CLIC energy
range. By constructing and operating the machine in a few energy stages the
luminosity performance over the full CLIC energy range can be maximized, without
compromising the re-use of equipment from the earlier stages at the higher
energies. At each stage the energy can be tuned down within a range of a factor
$\approx$ 3 with limited luminosity performance loss.

\begin{figure}[tb]
  \centering
  \includegraphics[width=0.75\textwidth,clip,trim=10 30 10 10]{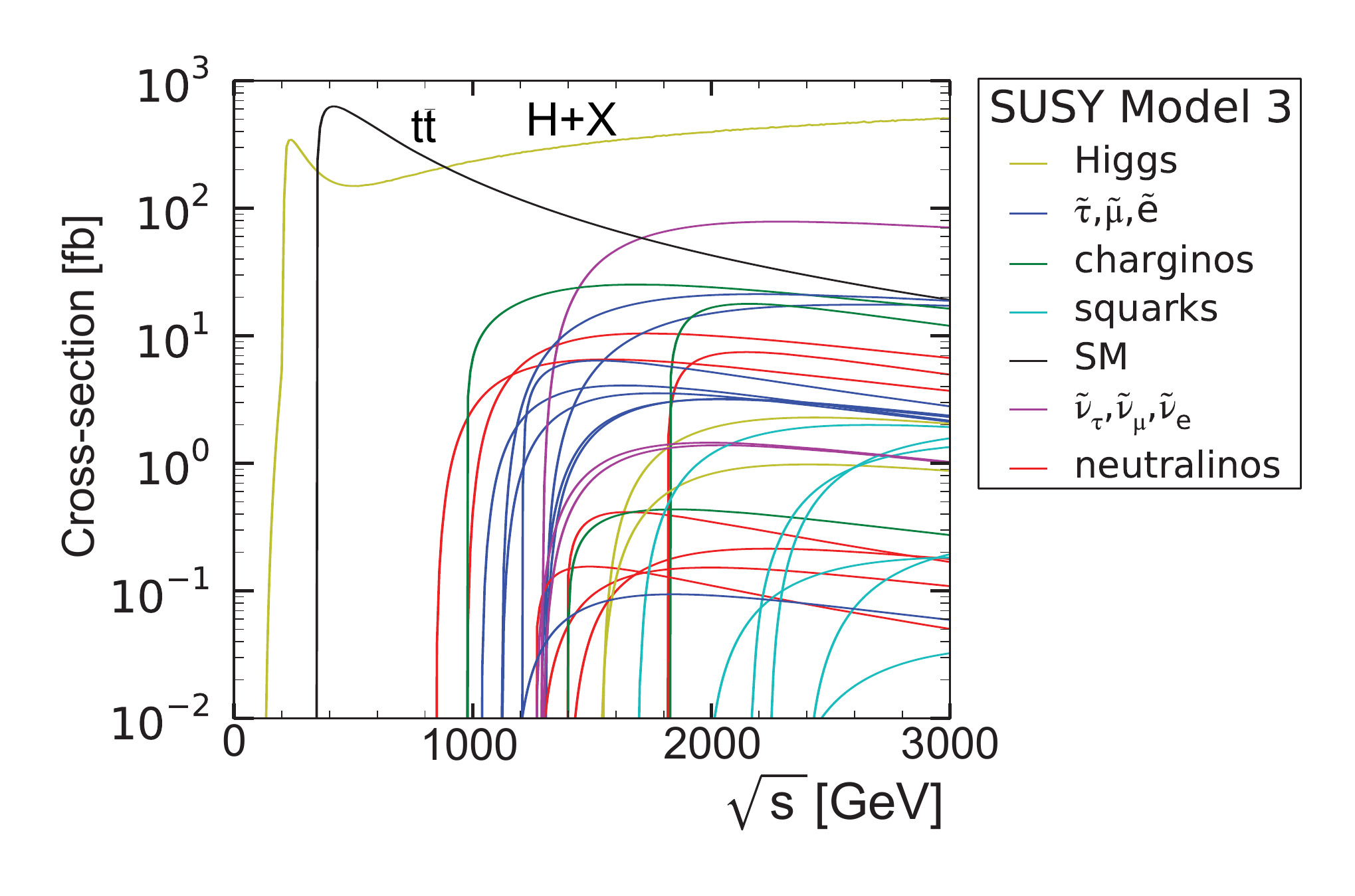}
  \caption{Higgs, \ttbar and SUSY production cross sections of \modelthree as a function of $\roots$. 
    Every line of a given color corresponds to the production cross
    section of one particle in the legend.}
  \label{fig:higgs_modelIII_xsec}
\end{figure}

Final choices for the CLIC energy stages will depend on the physics results from LHC running at 14~\TeV. For illustration, a possible staging scenario can be inferred from \autoref{fig:higgs_modelIII_xsec}, showing cross sections for Higgs and top production and for a possible SUSY scenario as a function of the center-of-mass energy. A first CLIC energy stage around 350~GeV to 375~GeV is very well motivated by Higgs and top physics. A next energy stage, around 
$\approx$ 1.4~\TeV, would give access to the detection and precise measurements of gauginos and sleptons in the example SUSY model (labelled \modelthree~\cite{CLICCDR_vol3}), while also providing additional Higgs precision measurements through vector-boson fusion, ttH production and double-Higgs production. A third stage at the highest CLIC energy of 3~\TeV would complement precision Higgs physics, \eg for decays with very small branching ratios, while also giving access to heavy Higgs partners and squarks in the example SUSY scenario. 

In the following sections, the CLIC physics capabilities with this staging scenario are elaborated in more detail. \autoref{sec:higgs-physics} describes the potential for precision Higgs physics and its impact on our knowledge of the underlying Electroweak Symmetry Breaking (EWSB) process and of New Physics scenarios. \autoref{sec:top-physics} gives an overview of top physics at CLIC\@. The section puts emphasis on top mass measurements, and illustrates other examples of top physics at CLIC, which will be studied in more detail in the near future. \autoref{sec:bsm-searches} illustrates the CLIC potential for physics searches beyond the standard model. Precision measurements are mostly already included in the Higgs, top and BSM sections, and are summarized together in \autoref{sec:prec-meas}. Overall conclusions and an outlook are presented in \autoref{sec:summary-conclusions}.

%%% Local Variables: 
%%% TeX-PDF-mode: t 
%%% TeX-master: "../clicsnowmass"
%%% End: 

\bibliocommand

%%% Local Variables: 
%%% TeX-PDF-mode: t 
%%% TeX-master: "../clicsnowmass"
%%% End: 

%\include{machine_detector/machine_detector}
\section{Higgs Physics at CLIC}
\label{sec:higgs-physics}

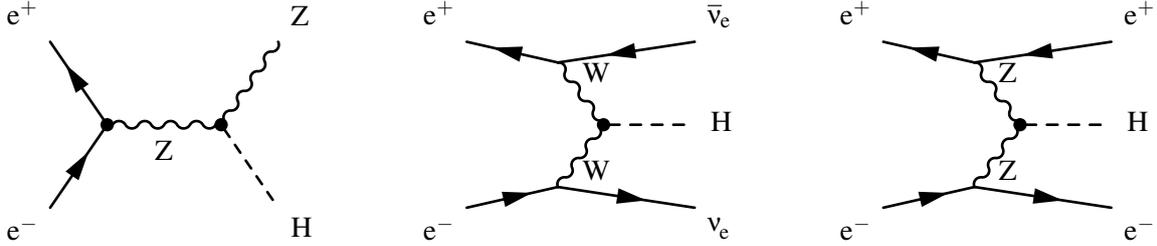
\begin{figure}[htb]
\unitlength = 1mm
\vspace{6mm}
\centering
\begin{fmffile}{higgs/eezh}
\begin{fmfgraph*}(30,22)
\fmfstraight
\fmfleft{i1,i2}
\fmfright{o1,o2}
\fmflabel{$\Pem$}{i1}
\fmflabel{$\Pep$}{i2}
\fmflabel{$\PZ$}{o2}
\fmflabel{$\PH$}{o1}
\fmf{photon,tension=1.0,label=$\PZ$}{v1,v2}
\fmf{fermion,tension=1.0}{i1,v1,i2}
\fmf{photon,tension=1.0}{o2,v2}
\fmf{dashes,tension=1.0}{v2,o1}
\fmfdot{v1}
\fmfdot{v2}
\end{fmfgraph*}
\end{fmffile}
\hspace{20mm}
\begin{fmffile}{higgs/eevvh}
\begin{fmfgraph*}(30,22)
\fmfstraight
\fmfleft{i1,i2}
\fmfright{o1,oh,o2}
\fmflabel{$\Pem$}{i1}
\fmflabel{$\Pep$}{i2}
\fmflabel{$\PAGne$}{o2}
\fmflabel{$\PH$}{oh}
\fmflabel{$\PGne$}{o1}
\fmf{fermion, tension=2.0}{i1,v1}
\fmf{fermion, tension=1.0}{v1,o1}
\fmf{fermion, tension=1.0}{o2,v2}
\fmf{fermion, tension=2.0}{v2,i2}
\fmf{photon, lab.side=right,lab.dist=1.5,label=$\PW$,tension=1.0}{v1,vh}
\fmf{photon, lab.side=right, lab.dist=1.5,label=$\PW$,tension=1.0}{vh,v2}
\fmf{dashes, tension=1.0}{vh,oh}
\fmfdot{vh}
\end{fmfgraph*}
\end{fmffile}
\hspace{20mm}
\begin{fmffile}{higgs/eeeeh}
\begin{fmfgraph*}(30,22)
\fmfstraight
\fmfleft{i1,i2}
\fmfright{o1,oh,o2}
\fmflabel{$\Pem$}{i1}
\fmflabel{$\Pep$}{i2}
\fmflabel{$\Pep$}{o2}
\fmflabel{$\PH$}{oh}
\fmflabel{$\Pem$}{o1}
\fmf{fermion, tension=2.0}{i1,v1}
\fmf{fermion, tension=1.0}{v1,o1}
\fmf{fermion, tension=1.0}{o2,v2}
\fmf{fermion, tension=2.0}{v2,i2}
\fmf{photon, lab.side=right,lab.dist=1.5,label=$\PZ$,tension=1.0}{v1,vh}
\fmf{photon, lab.side=right, lab.dist=1.5,label=$\PZ$,tension=1.0}{vh,v2}
\fmf{dashes, tension=1.0}{vh,oh}
\fmfdot{vh}
\end{fmfgraph*}
\end{fmffile}
\vspace{5mm}
\caption{The three highest cross section Higgs production processes at CLIC\@. At $\roots \approx 350~\GeV$, the Higgs-strahlung process dominates. Above $\roots \approx 500~\GeV$, the $\PW\PW$ vector boson fusion process  \mbox{$\Pep\Pem\to\PH\PGne\PAGne$} is dominant, with the cross section for the $\PZ\PZ$ process being about one order of magnitude lower. 
\label{fig:higgs:eezh}}
\end{figure}

A high-energy $\Pep\Pem$ collider such as CLIC would provide a clean environment to study the properties of the Higgs boson with 
very high precision.
The different CLIC energy stages will enable measurements of the properties of a 125~GeV Standard Model like Higgs boson through a number 
of different production processes, where those with the highest cross sections are indicated in \autoref{fig:higgs:eezh}. The sensitivities to 
Higgs physics at CLIC have been studied in the context of the staged scenario described previously. The results from these studies
have been obtained using detailed \textsc{Geant4} detector simulations, full reconstruction with the dominant background from $\gghadrons$ overlaid and inclusion of all relevant Standard Model background processes.

\begin{figure}[hbt]
\unitlength = 1mm 
\vspace{5mm}
\centering
         \begin{fmffile}{higgs/eetth} 
     \begin{fmfgraph*}(30,22)
        \fmfstraight
         \fmfleft{i1,i2}  
         \fmfright{o1,oh,o2}
         \fmflabel{$\Pem$}{i1}
        \fmflabel{$\Pep$}{i2}  
         \fmflabel{$\PQt$}{o2} 
         \fmflabel{$\PH$}{oh}
         \fmflabel{$\PAQt$}{o1}
          \fmf{photon,label=$\PZ$,tension=2.0}{v1,v2}
          \fmf{fermion,tension=1.0}{i1,v1,i2}
           \fmf{phantom,tension=1.0}{o1,v2,o2}
            \fmffreeze
           \fmf{fermion,tension=1.0}{o1,v2}
           \fmf{fermion,tension=1.0}{v2,vh,o2}
           \fmf{dashes,tension=0.0}{vh,oh}
           \fmfdot{vh}
	   \end{fmfgraph*}
	   \end{fmffile}
	   \hspace{15mm} 
	 \begin{fmffile}{higgs/eevvhh}
        \begin{fmfgraph*}(30,22)
            \fmfleft{i1,i2}  
            \fmfright{o1,oh1,oh2,o2}
            \fmflabel{$\Pem$}{i1}
            \fmflabel{$\Pep$}{i2}  
            \fmflabel{$\PAGne$}{o2} 
            \fmflabel{$\PH$}{oh1}            
            \fmflabel{$\PH$}{oh2}
            \fmflabel{$\PGne$}{o1}
             \fmf{fermion, tension=2.0}{i1,v1}
              \fmf{fermion, tension=1.0}{v1,o1}
             \fmf{fermion, tension=1.0}{o2,v2}
                          \fmf{fermion, tension=2.0}{v2,i2}
             \fmf{photon,  label=$\PW$,label.dist=1.5,tension=1.0}{v1,vh}
             \fmf{photon,  label=$\PW$,label.dist=1.5,tension=1.0}{v2,vh}
             \fmf{dashes, label=$\PH$, tension=2.0}{vh,vh1}
             \fmf{dashes,  tension=1.0}{oh1,vh1,oh2}
             \fmfdot{vh1}
        \end{fmfgraph*}
    \end{fmffile}
    \hspace{15mm}
	 \begin{fmffile}{higgs/ghhww}
        \begin{fmfgraph*}(30,22)
            \fmfleft{i1,i2}  
            \fmfright{o1,oh1,oh2,o2}
            \fmflabel{$\Pem$}{i1}
            \fmflabel{$\Pep$}{i2}  
            \fmflabel{$\PAGne$}{o2} 
            \fmflabel{$\PH$}{oh1}            
            \fmflabel{$\PH$}{oh2}
            \fmflabel{$\PGne$}{o1}
             \fmf{fermion, tension=2.0}{i1,v1}
              \fmf{fermion, tension=1.0}{v1,o1}
             \fmf{fermion, tension=1.0}{o2,v2}
              \fmf{fermion, tension=2.0}{v2,i2}
             \fmf{photon, label=$\PW$,label.dist=1.5,  tension=1.0}{v1,vh}
                          \fmf{photon, label=$\PW$,label.dist=1.5, tension=1.0}{v2,vh}
             \fmf{dashes,  tension=1.0}{oh1,vh,oh2}
             \fmfdot{vh}
        \end{fmfgraph*}
    \end{fmffile}
    \vspace{5mm}
\caption{The main processes at CLIC involving the  top-quark Yukawa coupling $g_{\PH\PQt\PQt}$, the Higgs boson trilinear self-coupling $\lambda$ and the quartic
              coupling $g_{\PH\PH\PW\PW}$.
 \label{fig:higgs:lambda}}
\end{figure}
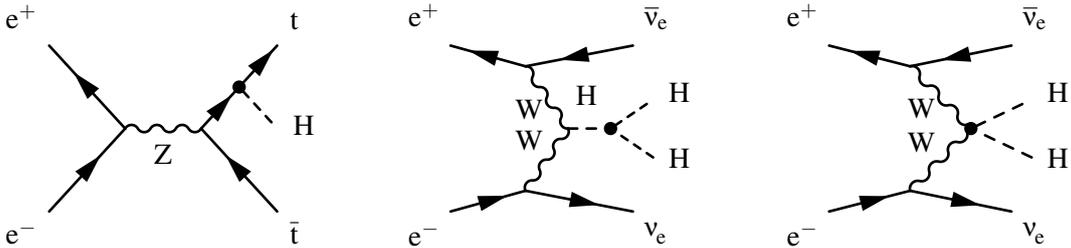

In the initial stage of CLIC operation at $\roots \approx 350~\GeV$, the Higgs-strahlung process $\Pep\Pem\to\PZ\PH$ dominates,
allowing a precise model-independent measurement of the coupling of the Higgs boson to the $\PZ$ and providing precise 
measurements of the Higgs boson branching ratios to a number of final states.
In the higher energy stages of CLIC operation (1.4~\TeV and 3.0~\TeV), large samples of Higgs bosons will be produced primarily through the
vector boson fusion process, $\Pep\Pem\to\PH\PGne\PAGne$. These large data samples will allow very precise ${\cal{O}}(1\%)$ 
measurements of the couplings of the Higgs boson to both fermions and the gauge bosons.
In addition to the main Higgs production processes, rarer processes such as
$\Pep\Pem\to\PQt\PAQt\PH$ and $\Pep\Pem\to\PH\PH\PGne\PAGne$, shown in \autoref{fig:higgs:lambda}, 
provide access to the top quark Yukawa coupling and the Higgs trilinear self-coupling as determined by the parameter $\lambda$ in the Higgs potential.

\begin{figure}[t]
  \centering
 \includegraphics[width=\halfwidth]{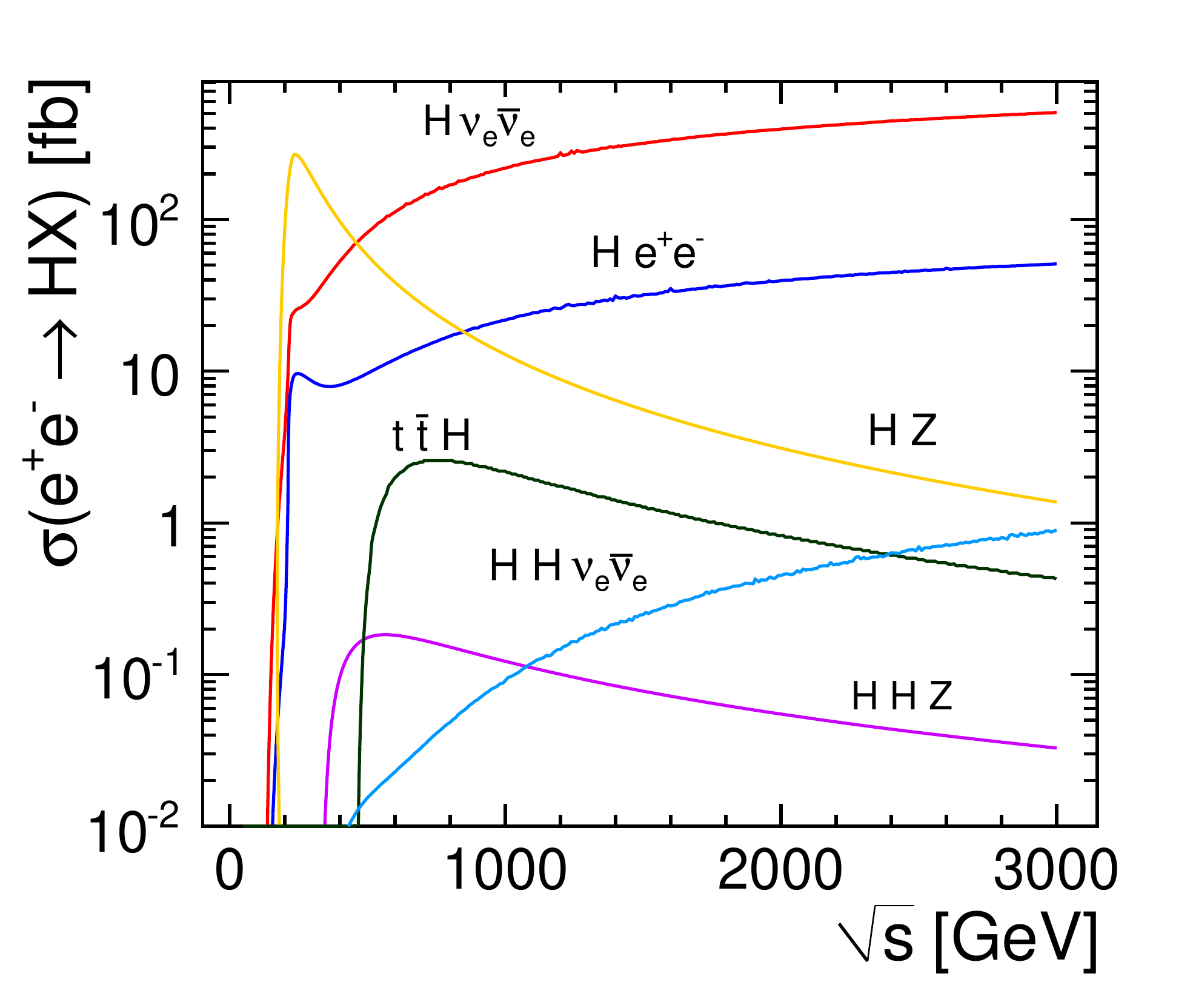}
 \caption{The center-of-mass dependencies of the cross sections for the main Higgs production processes at an $\Pep\Pem$ collider. The values shown correspond to unpolarized beams and do not include the effects of initial-stare radiation (ISR) or beamstrahlung.
 \label{fig:higgs:cross}}
\end{figure}

The raw Higgs production cross sections as a function of center-of-mass energy are shown in
\autoref{fig:higgs:cross}. The relatively large Higgs production cross sections, combined with the high integrated luminosities achievable at 
CLIC, result in large samples of Higgs bosons (far surpassing the number of $\PW$ bosons produced at LEP).
\autoref{tab:higgs:events} compares the expected number of $\PZ\PH$ and $\PH\PGne\PAGne$ events at the three
main center-of-mass energies considered in a CLIC energy staged scenario. The numbers of events include the effect of beamstrahlung, 
which results in a tail in the distribution of the effective center-of-mass energy $\rootsprime$.
Even at the lowest CLIC energies considered here, large and clean samples of Higgs boson decays can be accumulated. 
The $\PZ\PZ$ fusion process $\Pep\Pem\to\PH\Pep\Pem$ has a cross section that is approximately an order of magnitude smaller than the $\PW\PW$ fusion process, is also a significant source of Higgs bosons.

\begin{table}[tb]
  \centering
  \caption{The leading-order Higgs \emph{unpolarized} cross sections for the Higgs-strahlung, $\PW\PW$-fusion, and 
  $\PZ\PZ$-fusion processes    
for $\mH=125~\GeV$ at the three center-of-mass energies discussed in this document. The quoted cross sections 
include the effects of ISR but do not include the effects of beamstrahlung. Also listed are the numbers of 
expected events
\emph{including} the effects of beamstrahlung and ISR\@. The impact of beamstrahlung on the expected numbers of events is relatively
small, leading to an approximately 10\% reduction in the numbers of $\PH\PGne\PAGne$ events at $\roots>1~\TeV$. 
The cross sections and expected numbers do not account for the 
enhancements possible from polarized beams.
    \label{tab:higgs:events}}
  \begin{tabular}{lrrr}
    \toprule 
                                         & \tabt{350~GeV} & \tabt{1.4~TeV} & \tabt{3~TeV} \\ \midrule
    \LumiInt                             & 500~\fbinv     & 1500~\fbinv    & 2000~\fbinv  \\
%  No ISR, no beamstrahlung
%    $\sigma(\Pep\Pem\to\PZ\PH)$         & 129~fb         & 6.41~fb        & 1.37~fb      \\
%    $\sigma(\Pep\Pem\to\PH\PGne\PAGne)$ & 54.7~fb        & 294~fb         & 498~fb       \\
%    $\sigma(\Pep\Pem\to\PH\Pep\Pem)$    & 7.57~fb        & 29.4~fb        & 50.6~fb      \\
%  Including ISR but no beamstrahlung
%  $\sigma(\Pep\Pem\to\PZ\PH)$           & 134~fb         & 8.52~fb        & 2.03~fb      \\
%   $\sigma(\Pep\Pem\to\PH\PGne\PAGne)$  & 52.4~fb        & 279~fb         & 479~fb       \\
%   $\sigma(\Pep\Pem\to\PH\Pep\Pem)$     & 7.39~fb        & 27.9~fb        & 48.7~fb      \\
%  Including ISR but no beamstrahlung (rounded)
   $\sigma(\Pep\Pem\to\PZ\PH)$           & 134~fb         & 9~fb           & 2~fb         \\
    $\sigma(\Pep\Pem\to\PH\PGne\PAGne)$  & 52~fb          & 279~fb         & 479~fb       \\
    $\sigma(\Pep\Pem\to\PH\Pep\Pem)$     & 7~fb           & 28~fb          & 49~fb        \\
%  Including ISR and beamstrahlung
%   $\sigma(\Pep\Pem\to\PZ\PH)$          & 137~fb         & 13.6~fb        & 5.4~fb       \\
%   $\sigma(\Pep\Pem\to\PH\PGne\PAGne)$  & 52.1~fb        & 245~fb         & 416~fb       \\
%   $\sigma(\Pep\Pem\to\PH\Pep\Pem)$     & 7.40~fb        & 24.6~fb        & 42~fb        \\
    \# $\PZ\PH$ events                   & 68,000         & 20,000         & 11,000       \\
    \# $\PH\PGne\PAGne$ events           & 26,000         & 370,000        & 830,000      \\
    \# $\PH\Pep\Pem$ events              & 3,700          & 37,000         & 84,000       \\
    \bottomrule
  \end{tabular}
\end{table}

The measurement of the absolute coupling of the Higgs boson to the $\PZ$, which can be obtained from the recoil mass distribution in $\PH\PZ\to\PH\Pep\Pem$ and
$\PH\PZ\to\PH\PGmp\PGmm$ (see \autoref{sec:higgs:recoil}), plays a central role in the determination of the absolute Higgs couplings at a
linear collider. For this reason, it might seem surprising that no significant running is considered at $\roots= 250~\GeV$, which is close to the maximum of
the Higgs-strahlung cross section (see \autoref{fig:higgs:cross}). However, the reduction in cross section is, in part, compensated by the increased
instantaneous luminosity achievable at a higher center-of-mass energy; the instantaneous luminosity is expected to approximately linearly scale with the center-of-mass energy.
For this reason the precision on the coupling $g_{\PH\PZ\PZ}$ at $350~\GeV$ is comparable to that achievable at $250~\GeV$ for the same period of
operation, as indicated in \autoref{tab:higgs:zh}. Furthermore,  for the majority of final states, the measurement
\mbox{$\sigma(\PH\PZ)\times BR(\PH\to X)$} would be slightly more precise at 
$\roots=350~\GeV$. Initial operation at $\roots\approx350~\GeV$ also allows access to Higgs production
through the $\PW\PW$ fusion process, providing a constraint on the Higgs coupling to the $\PW$ boson. 
In addition, operation at $\roots\approx350~\GeV$ enables detailed studies of the top-pair production process. For these reasons, $\roots \approx 350~\GeV$ is the preferred option for the 
first stage of CLIC operation and no running at $\roots\approx250~\GeV$ is currently foreseen.

\begin{table}[t]\centering
  \caption{Precision measurements of the Higgs coupling to the $\PZ$ at $\roots=250~\GeV$ and
    $\roots=350~\GeV$ based on full
    simulation studies with $\mH=120~\GeV$. Results from \cite{ildloi:2009} and follow-up studies \cite{Li2011}.
    The numbers assume that the accelerator, in this case the ILC, operates with $-80\%,+30\%$ electron and positron
    beam polarizations.
    \label{tab:higgs:zh}}
  \begin{tabular}{lccc}\toprule
    $\roots$                              & 250~GeV    & 350~GeV    & 350~GeV    \\ \midrule
    \LumiInt                              & 250~\fbinv & 350~\fbinv & 500~\fbinv \\
    $\Delta(\sigma)/\sigma$               & 3\%        & 3.7\%      & 3.1\%      \\
    $\Delta(g_{\PH\PZ\PZ})/g_{\PH\PZ\PZ}$   & {1.5\%}    & {1.9\%}   & {1.6\%}    \\\bottomrule
  \end{tabular}
\end{table}

%%%%%%%%%%%%%%%%%%%%%%%%%%%%%%%%%%%%%
\subsection{\boldmath Higgs Measurements at {$\boldmath\roots=350$} GeV}

\label{sec:higgs:recoil}

\begin{figure}[t]
\centering
 \includegraphics[width=0.50\linewidth]{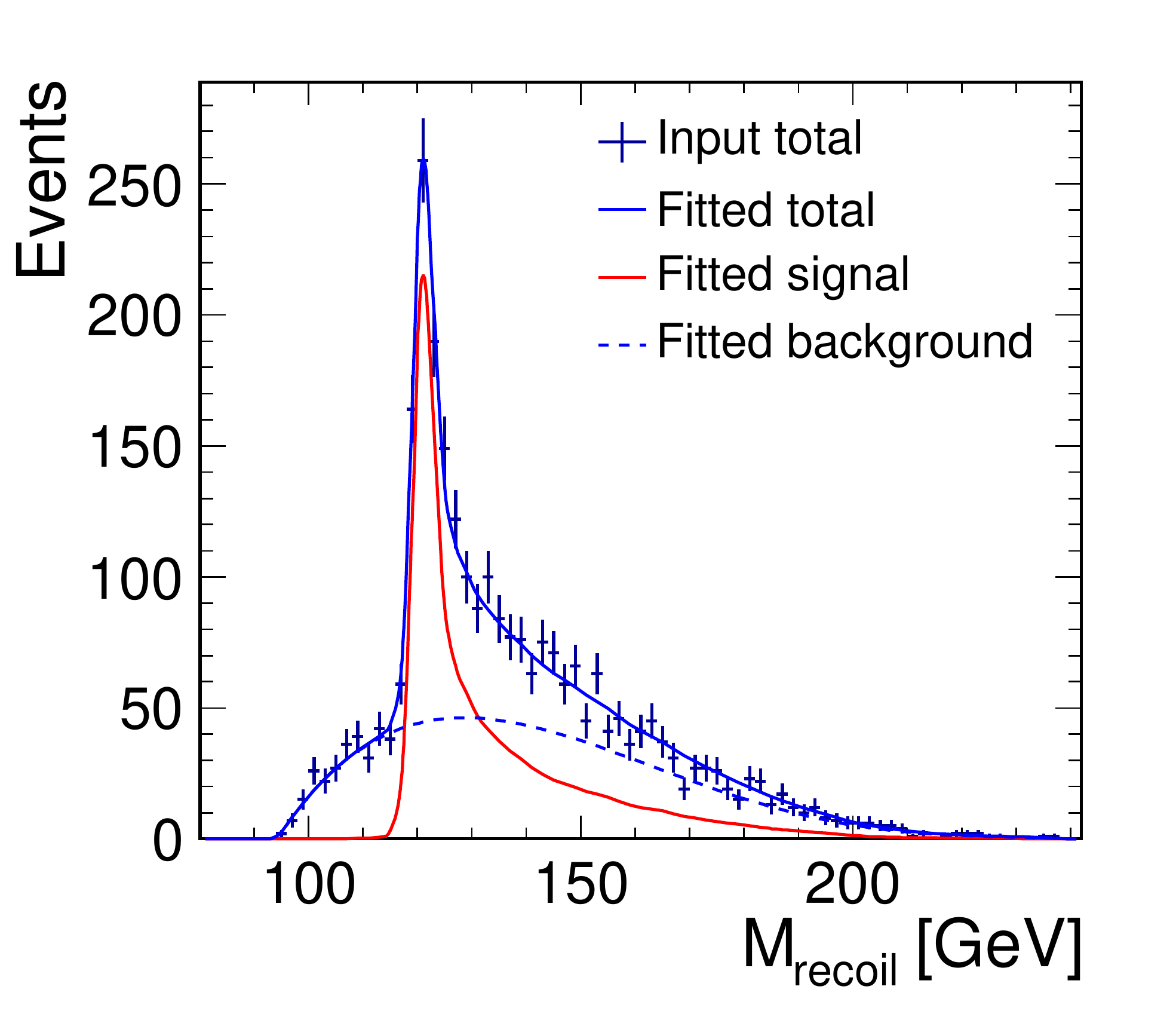}
 \caption{The recoil mass distribution for  $\Pep\Pem\to\PZ\PH\to\PGmp\PGmm\PH$ events with 
              $m_{\PH}=125~\GeV$ in the \clicild detector concept~\cite{CLICCDR_vol3}. The numbers of events correspond to 500~\fbinv at
              $\roots=350~\GeV$, and the error bars show the expected statistical uncertainties on the individual points.
 \label{fig:higgs:recoil}}
\end{figure}

The Higgs-strahlung process provides the opportunity to study the couplings of the Higgs boson in a
\textit{model-independent} manner. This is unique to an electron-positron collider. The
clean experimental environment, and the relatively low SM cross sections for
background processes,
allow  $\Pep\Pem\to\PZ\PH$ events to be selected based solely on the identification of two opposite
charged leptons with an invariant mass consistent with $\mZ$. The remainder of the event, 
the particles from the Higgs decay, is
not considered in the event selection. For example, \autoref{fig:higgs:recoil} shows the simulated invariant mass distribution of the 
system recoiling against identified $\PZ\to\PGmp\PGmm$ decays at CLIC for $\roots = 350~\GeV$.  A clear peak at the generated Higgs mass of $\mH=125~\GeV$ is 
observed. Because only the properties of the di-lepton system are used in the selection, this method provides an absolute measurement of the Higgs-strahlung cross section, 
regardless of the Higgs boson decay modes; it would be equally valid if the Higgs boson decayed to invisible final states.
Hence a model-independent measurement of the coupling $g_{\PH\PZ\PZ}$ can be made. With a dedicated analysis using also the hadronic decays of the $\PZ$, 
the sensitivity to invisible decay modes can be improved significantly. An $\epem$ linear 
collider provides unique sensitivity to invisible decay modes of the Higgs boson, extending down to a branching ratio into invisible states as low as $1\%$. 
For unpolarized beams, the study of the simulated $\PZ\PH$ recoil mass distributions for $\PZ\to\epem$ and $\PZ\to\mpmm$ decays at CLIC
operating at $\roots=350~\GeV$ gives a precision on the Higgs-strahlung cross section of approximately $4\%$~\cite{CLICCDR_vol3} for  $m_{\PH}=125~\GeV$. 

The recoil mass study provides an absolute measurement of the total $\PZ\PH$ production
cross section, and therefore the total number of Higgs bosons produced would be known with a statistical precision of $4\%$.
The systematic uncertainties from the knowledge of the integrated luminosity and event selection are expected to be significantly smaller.
Subsequently, by identifying the individual final states for different Higgs and $\PZ$ decay modes, measurements of the Higgs boson
branching fractions can be made. Because of the high flavor-tagging efficiencies~\cite{CLIC_PhysDet_CDR} achievable at CLIC, 
the $\PH\to\PQb\PAQb$ and $\PH\to\PQc\PAQc$ decays can be cleanly separated. Neglecting the Higgs decays into light quarks, 
one can also infer the branching ratio of $\PH\to\Pg\Pg$. \autoref{tab:higgs:brs} summarizes the branching 
fraction precisions achievable at  CLIC  operating at 350~\GeV. This table also lists the measurement sensitivities achievable at CLIC operating above 1~TeV. The values in \autoref{tab:higgs:brs} represent a snapshot
of the CLIC Higgs analyses as of September 2013.

\begin{table}[htp]\centering
  \caption{The precisions obtainable for the Higgs observables at CLIC for integrated luminosities of 500~\fbinv at $\roots=350~\GeV$,
    1.5~\abinv at $\roots=1.4~\TeV$, and  2.0~\abinv at $\roots=3.0~\TeV$. In all cases unpolarized beams have been assumed. 
    The majority of the results are from the full detector simulation and reconstruction including overlaid background from 
    $\gghadrons$. The numbers marked by `$*$' are preliminary and the
    numbers marked by `$\dagger$' are estimates; these will be updated when full simulation results are available.
    The `$-$' indicates that a measurement is not possible or relevant at this center-of-mass energy and `tbd' indicates that 
     no results or estimates are yet available.   
    For the branching ratios, the measurement precision refers to the expected statistical uncertainty on the 
    product of the relevant cross section
     and branching ratio; this is \emph{equivalent} to the  
     expected statistical uncertainty of the product of couplings divided by $\Gamma_{\PH}$.  For the measurements from
     the $\PQt\PAQt\PH$ and $\PH\PH\PGne\PAGne$ processes, the measurement precisions give the expected statistical uncertainties on the quantity or quantities 
     listed under the observable heading. 
     In the fits described in \autoref{sec:higgs:fits}, event rates multiplied by a factor 1.8 (see \autoref{tab:higgs:polarization}) were assumed 
     for measurements of Higgs production in WW-fusion above 1~\TeV to simulate the effect of $-80\%$ electron polarization. This approach is conservative, 
     because all backgrounds including those from $s$-channel processes were scaled by the same amount as the signals.
    \label{tab:higgs:brs}}
  \begin{tabular}{lllccc}\toprule
                        &                                                           &                              & \tabttt{Statistical precision}                        \\\cmidrule(l){4-6}
        \tabt{Channel}  & \tabt{Measurement}                                        & \tabt{Observable}            & 350~GeV         & 1.4~TeV         & 3.0~TeV           \\ 
                        &                                                           &                              & 500~\fbinv      & 1.5~\abinv      & 2.0~\abinv        \\ \midrule
    $\PZ\PH$            & Recoil mass distribution                                  & $\mH$                        & $120~\MeV$      & $-$             & $-$               \\
    $\PZ\PH$            & $\sigma(\PH\PZ)\times BR(\PH\to\text{invisible})$         & $\Gamma_\text{inv}$          & tbd             & $-$             & $-$               \\
    $\PZ\PH$            & $\PH\to\PQb\PAQb$ mass distribution                       & $\mH$                        & tbd             & $-$             & $-$               \\
    $\PH\PGne\PAGne$    & $\PH\to\PQb\PAQb$ mass distribution                       & $\mH$                        & $-$             & $40~\MeV^*$     & $33~\MeV^*$       \\ \midrule
    $\PZ\PH$            & $\sigma(\PH\PZ)\times BR(\PZ\to\ell^+\ell^-)$             & $\gHZZ^{2}$                  & $4.2\%$         & $-$             & $-$               \\
    $\PZ\PH$            & $\sigma(\PH\PZ)\times BR(\PH\to\PQb\PAQb)$                & $\gHZZ^{2}\gHbb^{2}/\GH$     & $1\%^{\dagger}$ & $-$             & $-$               \\
    $\PZ\PH$            & $\sigma(\PH\PZ)\times BR(\PH\to\PQc\PAQc)$                & $\gHZZ^{2}\gHcc^2/\GH$       & $5\%^{\dagger}$ & $-$             & $-$               \\
    $\PZ\PH$            & $\sigma(\PH\PZ)\times BR(\PH\to\Pg\Pg)$                   &                              & $6\%^{\dagger}$ & $-$             & $-$               \\
    $\PZ\PH$            & $\sigma(\PH\PZ)\times BR(\PH\to\PGtp\PGtm)$               & $\gHZZ^{2}\gHTauTau^{2}/\GH$ & $5.7\%$         & $-$             & $-$               \\
    $\PZ\PH$            & $\sigma(\PH\PZ)\times BR(\PH\to\PW\PW^*)$                 & $\gHZZ^{2}\gHWW^{2}/\GH$     & $2\%^{\dagger}$ & $-$             & $-$               \\
    $\PZ\PH$            & $\sigma(\PH\PZ)\times BR(\PH\to\PZ\PZ^*)$                 & $\gHZZ^{2}\gHZZ^{2}/\GH$     & tbd             & $-$             & $-$               \\
    $\PH\PGne\PAGne$    & $\sigma(\PH\PGne\PAGne)\times BR(\PH\to\PQb\PAQb)$        & $\gHWW^{2}\gHbb^{2}/\GH$     & $3\%^{\dagger}$ & $0.3\%$         & $0.2\%$           \\
    $\PH\PGne\PAGne$    & $\sigma(\PH\PGne\PAGne)\times BR(\PH\to\PQc\PAQc)$        & $\gHWW^{2}\gHcc^{2}/\GH$     & $-$             & $2.9\%$         & $2.7\%$           \\
    $\PH\PGne\PAGne$    & $\sigma(\PH\PGne\PAGne)\times BR(\PH\to\Pg\Pg)$           &                              & $-$             & $1.8\%$         & $1.8\%$           \\
    $\PH\PGne\PAGne$    & $\sigma(\PH\PGne\PAGne)\times BR(\PH\to\PGtp\PGtm)$       & $\gHWW^{2}\gHTauTau^{2}/\GH$ & $-$             & $3.7\%$         & tbd               \\
    $\PH\PGne\PAGne$    & $\sigma(\PH\PGne\PAGne)\times BR(\PH\to\PGmp\PGmm)$       & $\gHWW^{2}\gHMuMu^{2}/\GH$   & $-$             & $29\%^*$        & $16\%$            \\
    $\PH\PGne\PAGne$    & $\sigma(\PH\PGne\PAGne)\times BR(\PH\to\upgamma\upgamma)$ &                              & $-$             & $15\%^*$        & tbd               \\
    $\PH\PGne\PAGne$    & $\sigma(\PH\PGne\PAGne)\times BR(\PH\to\PZ\upgamma)$      &                              & $-$             & tbd             & tbd               \\
    $\PH\PGne\PAGne$    & $\sigma(\PH\PGne\PAGne)\times BR(\PH\to\PW\PW^*)$         & $\gHWW^{4}/\GH$              & tbd             & $1.1\%^*$         & $0.8\%^*$         \\
    $\PH\PGne\PAGne$    & $\sigma(\PH\PGne\PAGne)\times BR(\PH\to\PZ\PZ^*)$         & $\gHWW^{2}\gHZZ^{2}/\GH$     & $-$             & $3\%^{\dagger}$ & $2\%^{\dagger}$   \\
    $\PH\Pep\Pem$       & $\sigma(\PH\Pep\Pem)\times BR(\PH\to\PQb\PAQb)$           & $\gHZZ^{2}\gHbb^{2}/\GH$     & $-$             & $1\%^{\dagger}$ & $0.7\%^{\dagger}$ \\ \midrule
    $\PQt\PAQt\PH$      & $\sigma(\PQt\PAQt\PH)\times BR(\PH\to\PQb\PAQb)$          & $\gHtt^{2}\gHbb^{2}/\GH$     & $-$             & $8\%$         & tbd               \\
    $\PH\PH\PGne\PAGne$ & $\sigma(\PH\PH\PGne\PAGne)$                               & $g_{\PH\PH\PW\PW}$           & $-$             & $7\%^*$         & $3\%^*$           \\
    $\PH\PH\PGne\PAGne$ & $\sigma(\PH\PH\PGne\PAGne)$                               & $\lambda$                    & $-$             & $28\%$          & $16\%$            \\
    $\PH\PH\PGne\PAGne$ & with $-80\%$ $\Pem$ polarization                             & $\lambda$                    & $-$             & $21\%$          & $12\%$            \\\bottomrule
  \end{tabular}
\end{table}

%%%%%%%%%%%%%%%%%%%%%%%%%%%%%%%%%%%%
\subsection{\boldmath Higgs Measurements at $\roots > 1$\,TeV}

The large samples of events from both $\PW\PW$ and $\PZ\PZ$ fusion processes that could be accumulated above a center-of-mass energy of
1~TeV would lead to a measurement of the relative couplings of the Higgs boson to the $\PW$  and $\PZ$ bosons at the ${\cal{O}}(1\%)$ level,
providing a strong test of the Standard Model prediction for \mbox{$g_{\PH\PW\PW} / g_{\PH\PZ\PZ}  =  \cos^2 \theta_\mathrm{W}$}.

In addition, the ability for clean flavor tagging combined with the large samples of $\PW\PW$ fusion events
allows the production rate of $\Pep\Pem\to\PH\PGne\PAGne\to\PQb\PAQb\PGne\PAGne$ to be determined with a statistical 
precision of much better than 1\%. In general, the Higgs production cross section multiplied by the appropriate Higgs boson 
decay branching ratios can be measured more precisely at high energies, as can be seen from \autoref{tab:higgs:brs}. 
The uncertainties of the Higgs boson couplings to fermions and gauge bosons can be obtained by combining the high-energy CLIC results with those 
from the Higgs-strahlung process at $\roots=350~\GeV$. Furthermore, the high statistics samples 
from the  $\Pep\Pem\to\PH\PGne\PAGne$ alone would provide precise measurements of {\it relative} Higgs branching ratios.
For example,  CLIC operating at 3~TeV would yield a statistical precision of 1.5\% on the ratio $g_{\PH\PQc\PQc}/g_{\PH\PQb\PQb}$, providing a
direct comparison of the Standard Model coupling predictions for up-type (charge $+2/3$) and down-type (charge $-1/3$) quarks. 

Finally, CLIC operation at $\roots = 1.4~\TeV$ and above enables a determination of the top Yukawa coupling from the process 
$\Pep\Pem\to\PQt\PAQt\PH\to\PQb\PW^+\PAQb\PW^-\PH$. This process has been 
studied for the cases where the Higgs boson decays to $\PQb\PAQb$ and $\PWp\PWm$  decays 
either fully hadronically ($\text{q}\overline{\text{q}}\text{q}\overline{\text{q}}$) or semi-leptonically 
($\text{q}\overline{\text{q}}\ell\PGn$). Despite the complex final states of six or eight jets, 
it has been shown that the top Yukawa coupling can be measured with a precision of 4\%.

%%%%%%%%%%%%%%%%%%%%%%%%%%%%%%%%%%%%%
\subsubsection{Impact of Beam Polarization}

\label{sec:higgs:polarization}
To date, all CLIC Higgs physics studies were performed assuming unpolarized $\Pep$ and $\Pem$ beams. 
However, for CLIC the baseline electron polarization is $\pm80\%$ and there is the possibility of positron polarization at a lower level. 
For an electron polarization of $P_-$ and positron polarization of $P_+$, the relative fractions of collisions in the different polarization states are
\begin{align*}
  \PemR\PepR\ : \ \mbox{$\frac{1}{4}$}(1+P_-)(1+P_+)\,  & \quad\quad \ \, \PemR\PepL\ : \  \mbox{$\frac{1}{4}$}(1+P_-)(1-P_+)\, \\
  \PemL\PepR\ : \ \mbox{$\frac{1}{4}$}(1-P_-)(1+P_+)\,  & \ \ \text{and} \ \ \PemL\PepL\ : \  \mbox{$\frac{1}{4}$}(1-P_-)(1-P_+)\,.
\end{align*} 
Consequently, the $s$-channel $\Pep\Pem\to\PZ\PH$ process and, in particular, the $t$-channel $\Pep\Pem\to\PH\PGne\PAGne$ process can be enhanced by beam polarization, as indicated in \autoref{tab:higgs:polarization}. The chiral nature of the weak coupling to fermions results in significant possible enhancements in the $\PW\PW$-fusion Higgs production mechanism. The results listed in \autoref{tab:higgs:brs} assume no beam polarization, although significant improvements in precision could be obtained if one were to assume $-80\%$ (left-handed) electron beam polarization (and possible additional positron polarization). For Higgs production in the $\Pep\Pem\to\PH\Pep\Pem$ process the cross-section dependence on the polarization is only moderate, e.g. an enhancement factor of 1.17 is found for $-80\%$ electron polarization at 1.4 and 3~\TeV. In practice, the balance between operation with different beam polarizations will depend on the CLIC physics program taken as a whole. 

\begin{table}[tb]\centering
\caption{Increases in the event rates for the $s$-channel $\Pep\Pem\to\PZ\PH$ process and for $\Pep\Pem\to\PH\PGne\PAGne$ at 1.4 and 3~\TeV (dominated by the $t$-channel $\PW\PW$-fusion pocesses) for three example beam polarizations. 
\label{tab:higgs:polarization}}
  \begin{tabular}{ccc}\toprule
    Polarization                & \tabtt{Enhancement factor}                        \\ \cmidrule(l){2-3}
    $P(\Pem):P(\Pep)$           & $\Pep\Pem\to\PZ\PH$ & $\Pep\Pem\to\PH\PGne\PAGne$ \\ \midrule
    unpolarized                 & 1.00                & 1.00                        \\
    $-80\%\,:\phantom{+3}\,0\%$ & 1.18                & 1.80                        \\
    $-80\%\,:\,+30\%$           & 1.48                & 2.34                        \\\bottomrule
  \end{tabular}
\end{table}

\subsection{Higgs Self-Coupling}

In the SM, the Higgs boson originates from a doublet of complex scalar fields described by the potential
\begin{equation*}
       V(\phi) = \mu^2\phi^\dagger\phi + \lambda(\phi^\dagger\phi)^2 \,.
\end{equation*}
After spontaneous symmetry breaking, this form of the potential gives rise to a trilinear Higgs self-coupling of strength proportional to $\lambda v$, where 
$v$ is the vacuum expectation value of the Higgs potential. The measurement of the strength of the Higgs self-coupling therefore provides direct access to 
the quartic potential coupling $\lambda$ assumed in the Higgs mechanism. This measurement is therefore
an essential part of experimentally establishing the Higgs mechanism as described by the Standard Model. For \mbox{$m_{\PH}=125~\GeV$}, the measurement of 
the Higgs boson self-coupling at the LHC will be extremely challenging even with 3000~\fbinv of data. 

At a linear collider, the trilinear Higgs coupling can be measured through the \mbox{$\Pep\Pem\to\PZ\PH\PH$} and 
$\Pep\Pem\to\PH\PH\PGne\PAGne$
processes. The achievable precision  has been studied for the $\Pep\Pem\to\PZ\PH\PH$ process at 
$\roots=500~\GeV$ in the context of the International Linear Collider (ILC), where the results show that a very large integrated luminosity is required~\cite{ILCPhysicsDBD}. 
For this reason, the most favorable channel for the measurement of the Higgs self-coupling is the $\Pep\Pem\to\PH\PH\PGne\PAGne$ process at 
$\roots\ge 1~\TeV$. Here the sensitivity increases with increasing center-of-mass energy. The $\lambda$ uncertainty is evaluated by measuring the uncertainty of the $\Pep\Pem\to\PH\PH\PGne\PAGne$ process cross section and relating this uncertainty to that of $\lambda$ via a conversion factor, which amounts to 1.20 and 1.54 at 1.4~\TeV and 3~\TeV, respectively. An alternative approach employs template fitting of the neural net classifier response distribution in order to obtain the uncertainty on $\lambda$ directly. The latter approach is preferred since it excludes potential dependence of uncertainty-related factors on the event selection. Results from a detailed study  
indicate that a precision of $28\%$ and $16\%$ on $\lambda$ can be achieved at CLIC operating respectively at $\roots=1.4~\TeV$ and 
$\roots=3~\TeV$, see \autoref{tab:higgs:brs}. These preliminary studies were performed assuming unpolarized beams. With $80\%$ left-handed 
polarized electrons and $30\%$ right-handed polarized positrons the 
signal cross section increases significantly and a measurement precision of $\approx10\%$ on 
$\lambda$ is attainable with CLIC operation at \mbox{$\roots=3~\TeV$}.

The $\Pep\Pem\to\PH\PH\PGne\PAGne$ process can proceed through several lowest-order Feynman diagrams. The diagram involving
the coupling $\lambda$ is shown in \autoref{fig:higgs:lambda} (center). The expected precision on $\lambda$ quoted above assumes that the 
contributions to the production cross section from other diagrams take their Standard Model values. However, the analysis of the
process $\Pep\Pem\to\PH\PH\PGne\PAGne$ can be interpreted differently, for example as measurement of the quartic coupling
at the $\PH\PH\PW\PW$ vertex, where the sensitivity comes from the Feynman diagram shown in \autoref{fig:higgs:lambda} (right).
A preliminary study indicates that measurement precisions of $7\%$ and $3\%$ on $g_{\PH\PH\PW\PW}$ can be achieved at CLIC operating respectively
at \mbox{$\roots=1.4~\TeV$} and \mbox{$\roots=3~\TeV$}.

\subsection{Higgs Boson Couplings and Total Decay Width}

\label{sec:higgs:fits}

In the previous sections, the Higgs boson measurements at CLIC were reviewed. In most cases the measurements correspond to the product of the 
Higgs production cross section with the relevant Higgs branching ratio. The absolute couplings of the Higgs boson can be determined using the total 
$\epem\to\PZ\PH$ cross section from the recoil mass distribution from $\PZ\to\Pem\Pep$ and, in particular,
$\PZ\to\PGmp\PGmm$,
measured at $\roots = 350~\GeV$. This allows the coupling of the Higgs boson to the $\PZ$ to be determined with a precision of about 2\% in a 
model-independent manner. Once the coupling to the $\PZ$ is known, the Higgs coupling to the $\PW$ can be determined from, for example, 
the ratios of Higgs-strahlung to $\PW\PW$ fusion cross sections,
\begin{equation*}
    \frac{\sigma(\epem\to\PZ\PH)\times BR(\PH\to\PQb\PAQb) }{ \sigma{(\epem\to\PGne\PAGne\PH)} \times BR(\PH\to\PQb\PAQb)  } \propto \left(\frac{g_{\PH\PZ\PZ}}{g_{\PH\PW\PW}} \right)^2 \,.
\end{equation*}  

In order to determine absolute measurements of the other Higgs couplings, the Higgs total decay width needs to be inferred from the data. 
For a Higgs boson mass of 125~\GeV, the total Higgs decay width in the SM ($\Gamma_{\PH}$) is less than 5 MeV and
cannot be measured directly.  However, given that the absolute 
couplings of the Higgs boson to the $\PZ$ and $\PW$ can be obtained as described above, the total decay width of the 
Higgs boson can be determined from $\PH\to\PW\PW^*$ or $\PH\to\PZ\PZ^*$ decays. For example, the measurement of the 
Higgs decay to $\PW\PW^*$ in the $\PW\PW$ fusion process determines
\begin{equation*}
                 \sigma(\PH\PGne\PAGne)\times BR(\PH\to\PW\PW^*)   \propto \frac{g^4_{\PH\PW\PW}}{\Gamma_{\PH}}\,,
\end{equation*}
and thus the total width can be determined utilizing the model-independent measurement of $g_{\PH\PW\PW}$. In practice, a fit would 
be performed to all of the experimental measurements involving the Higgs boson couplings. \autoref{tab:higgs:params} lists the results of such 
a fit \cite{LCD:2013-012}, applied using the expected statistical measurement precisions given in \autoref{tab:higgs:brs} (scaled to include the benefits of electron polarization).
Here it is assumed that CLIC operation above 1~TeV is primarily with $-80\%$ electron polarization to enhance the $\PW\PW$ fusion cross section. 
These \textit{preliminary fit results} do not yet include the constraints on the invisible width that can be obtained from the Higgs-strahlung process at $\roots=350~\GeV$,
and will be updated accordingly. The interpretation of the results listed in \autoref{tab:higgs:params} requires some care as the uncertainties on $\gHZZ$,
$\gHWW$, $\gHbb$, $\gHbb$, and $\gHTauTau$ are almost 100\% correlated and are fixed by the precision to which $\gHZZ$ can be determined; the uncertainties
on the ratios of these couplings are typically \textit{much smaller} than the uncertainties on the absolute values of the couplings. 

\begin{figure}[tb]
  \centering
 \includegraphics[width=0.95\linewidth]{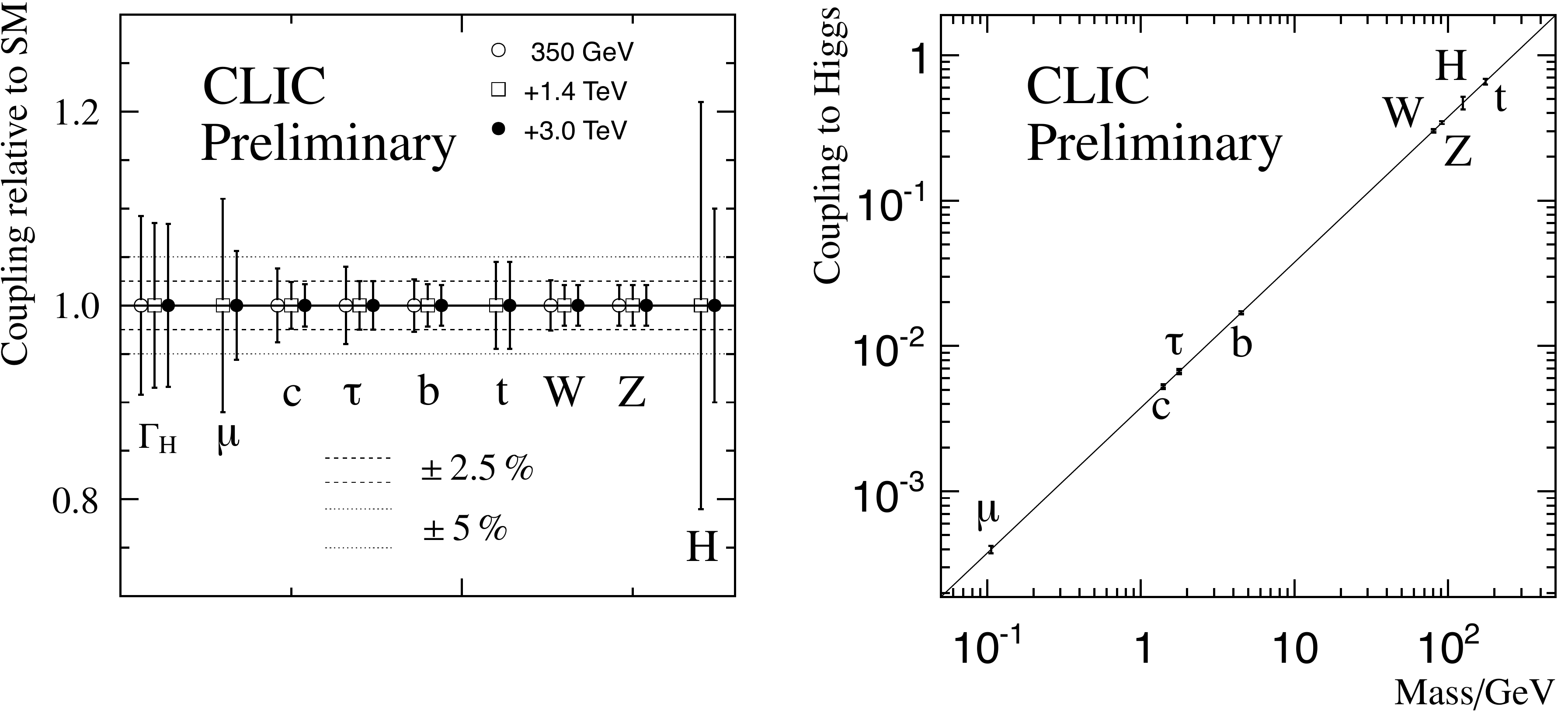}
\caption{An illustration of the typical precisions to which the relation between the Higgs couplings to
the masses of the particles can be tested at CLIC\@.
The left plot shows the precision achievable at CLIC after
                (i) an integrated luminosity of 500~\fbinv at $\roots=350~\GeV$,
                (ii) after an additional 1.5~\abinv at \mbox{$\roots=1.4~\TeV$}, and
                (iii) after a further 2.0~\abinv at $\roots=3.0~\TeV$.
 The listed values assume SM couplings. The right plot summarizes the precision of the test of the 
 prediction of the proportionality of the Higgs coupling to mass (after the three assumed stages of CLIC operation). 
 The sensitivities assume that operation above 1~TeV is primarily with $-80\%$ beam polarization. \ednote{This figure is just a place-holder}
 \label{fig:higgs:couplingrel}}
\end{figure}

Whilst the precise measurements of the Higgs couplings to gauge bosons, fermions and to itself
are of interest in their own right, they will be crucial for testing the fundamental prediction of the Higgs mechanism
that the Higgs coupling to different particles is proportional to their masses, as summarized in \autoref{fig:higgs:couplingrel}.

\begin{table}[btp]\centering
  \caption{The precision Higgs observables at CLIC after
    (i) an integrated luminosity of 500~\fbinv at $\roots=350~\GeV$,
    (ii) after an additional 1.5~\abinv at $\roots=1.4~\TeV$, and
    (iii) after a further 2.0~\abinv at $\roots=3.0~\TeV$.
    The results were obtained from the statistical measurement precisions quoted in \autoref{tab:higgs:brs}, scaled up assuming  
    that operation above 1~\TeV is primarily with $-80\%$ electron beam polarization.  The uncertainties are statistical only. The entries
    marked `tbd' indicate that results for $\roots = 3~\TeV$ have yet to be determined.
    \label{tab:higgs:params}}

  \begin{tabular}{lccc}\toprule
    \tabt{Parameter} & \tabttt{Measurement precision}             \\\cmidrule(l){2-4}
                     & 350~\GeV   & $+ 1.4~\TeV$    & $+ 3.0~\TeV$    \\
                     & 500~\fbinv & $+1.5~\abinv$ & $+2.0~\abinv$ \\\midrule 
    $\mH$              &   120~\MeV   &        30~\MeV    &     20~\MeV       \\
    $\GH$              &  9.2\%           &        8.5\%         &       8.4\%           \\
    $\lambda$        & $-$               &         21\%         &    10\%               \\
    $\gHZZ$          & $2.1\%$        & $2.1\%$            &   $2.1\%$           \\
    $\gHWW$        & $2.6\%$      & $2.1\%$             &   $2.1\%$           \\
    $\gHbb$           & $2.7\%$      & $2.2\%$            &    $2.1\%$           \\
    $\gHcc$           & $3.8\%$      & $2.4\%$            &     $2.2\%$           \\
    $\gHTauTau$   & $4.0\%$      & $2.5\%$            &      tbd                  \\
    $\gHMuMu$     & $-$             & $11\%$           &    $5.6\%$           \\
    $\gHtt$            & $-$              & $4.5\%$            &       tbd                  \\\bottomrule
  \end{tabular}
\end{table}

As an alternative to the model-independent approach discussed so far, a fit was performed 
using nine Higgs coupling scale factors defined as:
\begin{equation*}
\kappa_{i}^{2} = \frac{\Gamma_{i}}{\Gamma_{i}^{\rm SM}},
\end{equation*}
where $\Gamma_{i}$ and $\Gamma_{i}^{\rm SM}$ are the measurement and SM prediction for the partial width of an 
individual visible Higgs decay mode $i$  \cite{LCD:2013-012}. Here the total Higgs boson width is given by the sum of the nine considered partial decay widths. This 
is equivalent to the assumption of no invisible Higgs decays. In this scenario, the ratio of the 
fitted total Higgs width to its SM prediction can be calculated as:
\begin{equation*}
\Gamma_{\PH, {\rm model}} = \sum_{i}\kappa_{i}^{2} \cdot BR_{i}^{\rm SM}.
\end{equation*}
The SM expectations for the branching ratios assuming a Higgs mass of 126~\GeV, $BR_{i}^{\rm SM}$, were taken from~\cite{Dittmaier:2012vm}. 
Theoretical uncertainties on the SM predictions were neglected in the fit. \autoref{tab:higgs:kappas} shows the 
obtained \textit{preliminary} precisions on the Higgs coupling scale factors.

\begin{table}[btp]\centering
  \caption{The Higgs coupling scale factors at CLIC after
    (i) an integrated luminosity of 500~\fbinv at $\roots=350~\GeV$,
    (ii) after an additional 1.5~\abinv at $\roots=1.4~\TeV$, and
    (iii) after a further 2.0~\abinv at $\roots=3.0~\TeV$.
    The assumptions used for the determination of these parameters are explained in the text. 
    All results were obtained from the statistical measurement precisions quoted in \autoref{tab:higgs:brs}, scaled up assuming  
    that operation above 1~\TeV is primarily with $-80\%$ electron beam polarization. The uncertainties are statistical only. The entries
    marked `tbd' indicate that results for $\roots = 3~\TeV$ have yet to be determined.
    \label{tab:higgs:kappas}}

  \begin{tabular}{lccc}\toprule
    \tabt{Parameter} & \tabttt{Measurement precision}             \\\cmidrule(l){2-4}
                     & 350~\GeV   & $+ 1.4~\TeV$    & $+ 3.0~\TeV$    \\
                     & 500~\fbinv & $+1.5~\abinv$ & $+2.0~\abinv$ \\\midrule 
    $\Gamma_{\PH, \rm{model}}$ & 1.6\% & 0.29\% & 0.22\% \\
    $\kHZZ$ & $0.49\%$ & $0.33\%$ & $0.24\%$ \\
    $\kHWW$ & $1.5\%$ & $0.15\%$ & $0.11\%$ \\
    $\kHbb$ & $1.7\%$ & $0.33\%$ & $0.21\%$ \\
    $\kHcc$ & $3.1\%$ & $1.1\%$ & $0.75\%$ \\
    $\kHTauTau$ & $3.5\%$ & $1.4\%$ & tbd \\
    $\kHMuMu$ & $-$ & $11\%$ & $5.2\%$ \\
    $\kHtt$ & $-$ & $4.0\%$ & tbd \\
    $\kHGluGlu$ & $3.6\%$ & $0.79\%$ & $0.56\%$ \\
    $\kHGamGam$ & $-$ & $5.5\%$ & tbd \\
    \bottomrule
  \end{tabular}
\end{table}

%%%%%%%%%%%%%%%%%%%%%%%%%%%%%%%%%%
\subsection{Impact of the Precision Measurements of the Higgs Couplings}

The precise measurements of the Higgs boson properties at CLIC would provide
a powerful probe of the structure of the Higgs sector. The SM with a single Higgs doublet
is only one of the possibilities. The model-independent measurements at CLIC would
be crucial to distinguish between the different possible manifestations of the underlying physics.
In many extended Higgs theories the lightest Higgs scalar can have nearly identical properties
to the SM Higgs boson. In this decoupling limit, additional states of the Higgs sector are heavy and it may be difficult to detect them at the LHC\@. 
CLIC would provide sensitivity to new Higgs bosons beyond that achievable at the LHC, with states of masses up to 
essentially half the $\epem$ center-of-mass energy (see \autoref{sec:susy}) being directly detectable.  
Nevertheless, if these massive states were very heavy, and therefore beyond the reach of the LHC and even CLIC, 
precision measurements would be crucial in order to distinguish the simple Higgs sector of the SM from a more complicated scalar sector.

Deviations from the SM can arise from an extended structure of the Higgs sector, for
instance if there is more than one Higgs doublet. Another source of
possible deviations from the SM Higgs properties are loop effects from BSM
particles. The potential for probing the physics of EWSB is directly related to the sensitivity for verifying deviations from the SM\@.
% For example, in \autoref{fig:higgs:couplingbsm} the typical deviations from the SM predictions for a Two-Higgs-Doublet
% model are compared to the precision on the couplings achievable at the different energy stages of CLIC\@.
% In this example, the high-precision measurements at CLIC would clearly indicate the non-SM nature of the EWSB sector. 
% To observe such deviations at the $> 3 \sigma$ level, operation of CLIC at the higher energies is crucial.

Furthermore, small deviations from SM-like behavior can arise as a consequence of fundamentally different physics of EWSB\@. 
For example, if an additional fundamental scalar such as the radion mixes with the Higgs boson, the possibly small 
shifts in the branching ratios and overall decay width relative to the SM Higgs boson may only be discernible through
the high-precision and model-independent measurements of couplings available at a future linear collider. 

% \begin{figure}[bt]
% \centering
% \includegraphics[width=0.65\textwidth]{higgs/CLIC_2HDM.pdf}
% \caption{Typical deviations of the Higgs couplings to different
% particles from the SM predictions in a Two-Higgs-Doublet model. The CLIC
% precisions after the various energy stages are the same as in
% \autoref{fig:higgs:couplingrel}. \ednote{This figure is just a place-holder}
% \label{fig:higgs:couplingbsm}}
% \end{figure}

%%%%%%%%%%%%%%%%%%%%%%%%%%%%%%%%%%%
\subsection{Higgs Boson Mass, Spin and CP Properties}

Detailed studies of the properties of the Higgs boson are possible at a high-luminosity $\epem$ linear collider. For example the Higgs boson mass
can be determined to better than 100~MeV at CLIC operating at $\roots=350~\GeV$ from either the $\PZ$ recoil mass distribution or from the direct reconstruction of
the decay products. At higher center-of-mass energies the large samples of $\PH\to\PQb\PAQb$ decays would allow the Higgs mass to be determined with a statistical precision of about $\pm30~\MeV$. At this stage the potential size of systematic uncertainties, such as the $\PQb$-quark jet-energy scale, have not been assessed yet, although potential systematic biases could be mitigated using $\PZ\to\PQb\PAQb$ decays.

CLIC also provides the possibility of detecting CP violation in the Higgs sector, where
\textit{a priori}  the observed Higgs state with $m_{\PH} = 125~\GeV$ can be an admixture of
CP even and CP odd states. The most general model-independent expression for the $\PH VV$  coupling can be written as
\begin{equation*}
g_{\PH \text{VV}} =  -{g}{\text{M}_\text{V}}\left[ \alpha g_{\mu\nu} +
\beta \left(\frac{(p \cdot q)\, g_{\mu\nu}}{\text{M}_\text{V}^2} - p_\nu q_\mu  \right)
+ i\, \gamma \frac{\epsilon_{\mu\nu\rho\sigma}p^\rho q^\sigma}{\text{M}_\text{V}^2}
\right],
\end{equation*}
where V represents either a $\PW$ or $\PZ$ boson, $p$ and $q$ are the four momenta of the two vector bosons, and $\epsilon_{\mu\nu\rho\sigma}$
is the totally antisymmetric tensor. For the SM Higgs boson, $\alpha=1$ and $\beta=\gamma=0$. In contrast, for a pure CP odd Higgs boson, 
$\alpha = \beta = 0$, and $\gamma$ is expected to be small. Dedicated studies of the CP properties of the light Higgs boson have not yet been performed 
in the context of the CLIC study. Nevertheless, such studies have been performed for a $\epem$ linear collider operating at 
$\roots=350~\GeV$.  For example, by utilizing the angular correlations in $\Pep\Pem \to \PH\PZ \to 4\mathrm{f}$ it was 
demonstrated that $\eta$, which parameterizes the mixing between a CP-even and a CP-odd Higgs state, 
could be measured with an accuracy of $3\%\text{ -- }4\%$~\cite{TeslaTDR}. As part of the future CLIC study, it is intended to extend these studies to CLIC operating above 1~TeV.

\bibliocommand

%%% Local Variables:
%%% mode: latex
%%% TeX-master: "../clicsnowmass"
%%% TeX-PDF-mode: t
%%% End: 

\section{Top Physics}
\label{sec:top-physics}
%% Page limit: 3

\subsection{Introduction}

As the heaviest elementary particle known to date, the top quark is of particular interest. It couples most strongly to the Higgs field, and due to its high mass also provides leading contributions in higher order corrections to many processes and may provide high sensitivity to physics beyond the SM\@. Together with the Higgs mass, the top mass is a key input to studies of the stability of the SM vacuum, which can be seen as an upper validity bound of the SM\@. With the precision of the Higgs mass provided by the LHC today, the uncertainty of the top mass is the leading uncertainty in this evaluation. Improvements in the measurement of the top quark mass, possible at a linear collider, will substantially reduce these uncertainties. At the same time, the precise measurement of couplings of the top quark to the Higgs and to gauge bosons will provide the possibilities for precision tests of SM predictions and a corresponding sensitivity to New Physics at higher scales.

\subsection{Top Quark Mass Measurements at CLIC}

An \epem  collider offers two complementary ways of measuring the top quark mass; by direct reconstruction of the invariant mass of the decay products, and through a scan of the top pair production threshold. While the former measurement can be performed at essentially arbitrary energies above threshold, the latter requires collider operations at several closely-spaced energies around the pair production threshold. The theoretical interpretation of the two measurements differ considerably. The invariant mass is interpreted by comparing the measured distribution with that predicted by MC simulations, and as such is obtained in the context of the used event generator and is thus not defined in a theoretically rigorous way. Progress has been made in the calculation of non-perturbative effects which affect the final state of \ttbar decays, but uncertainties remain substantial. The cross section evolution around threshold is calculated to higher orders in theoretically well-defined mass schemes, providing a clean interpretation of the top quark mass obtained from a threshold scan.

\begin{figure}
\centering
  \includegraphics[width=\halfwidth]{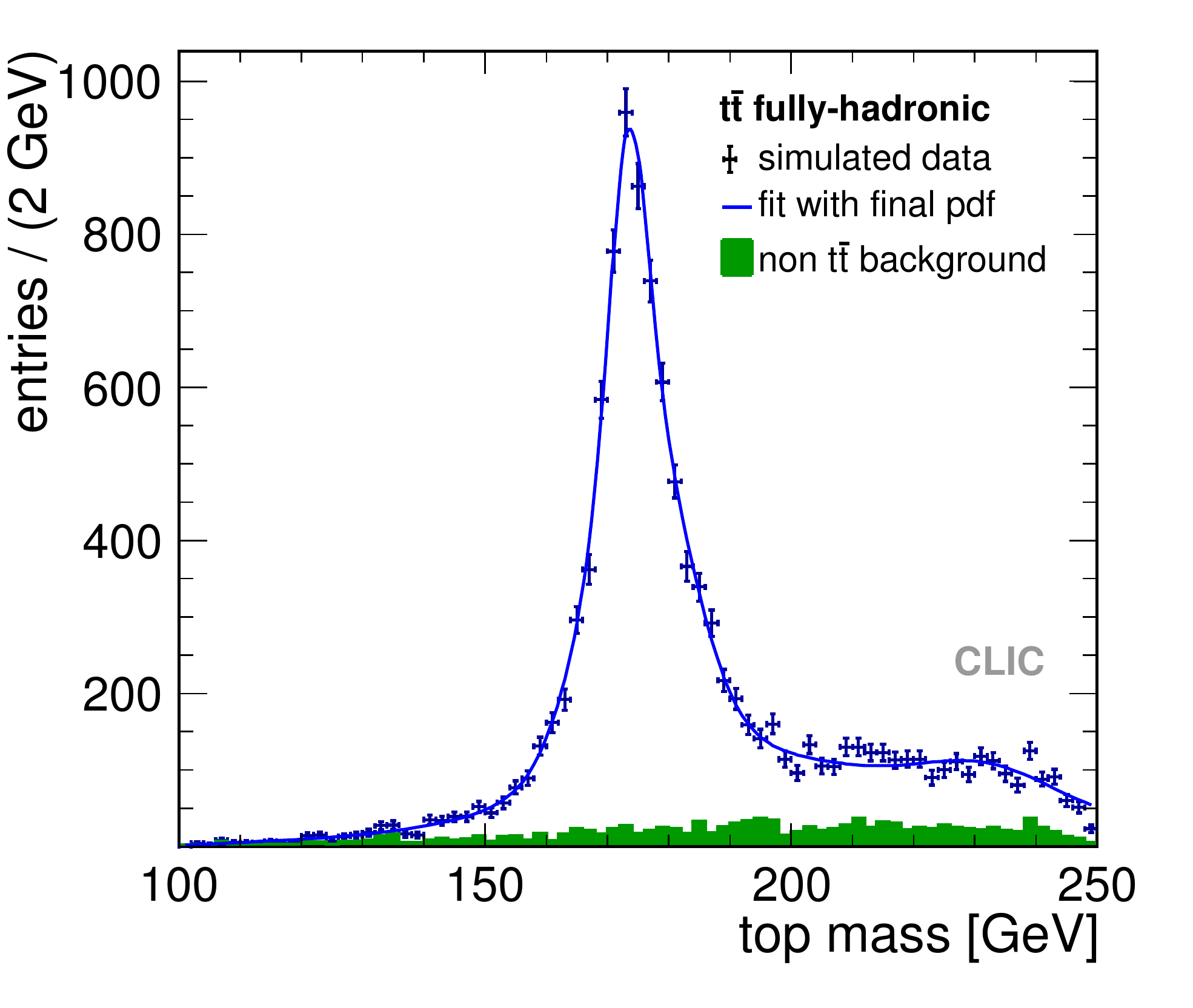}
  \includegraphics[width=\halfwidth]{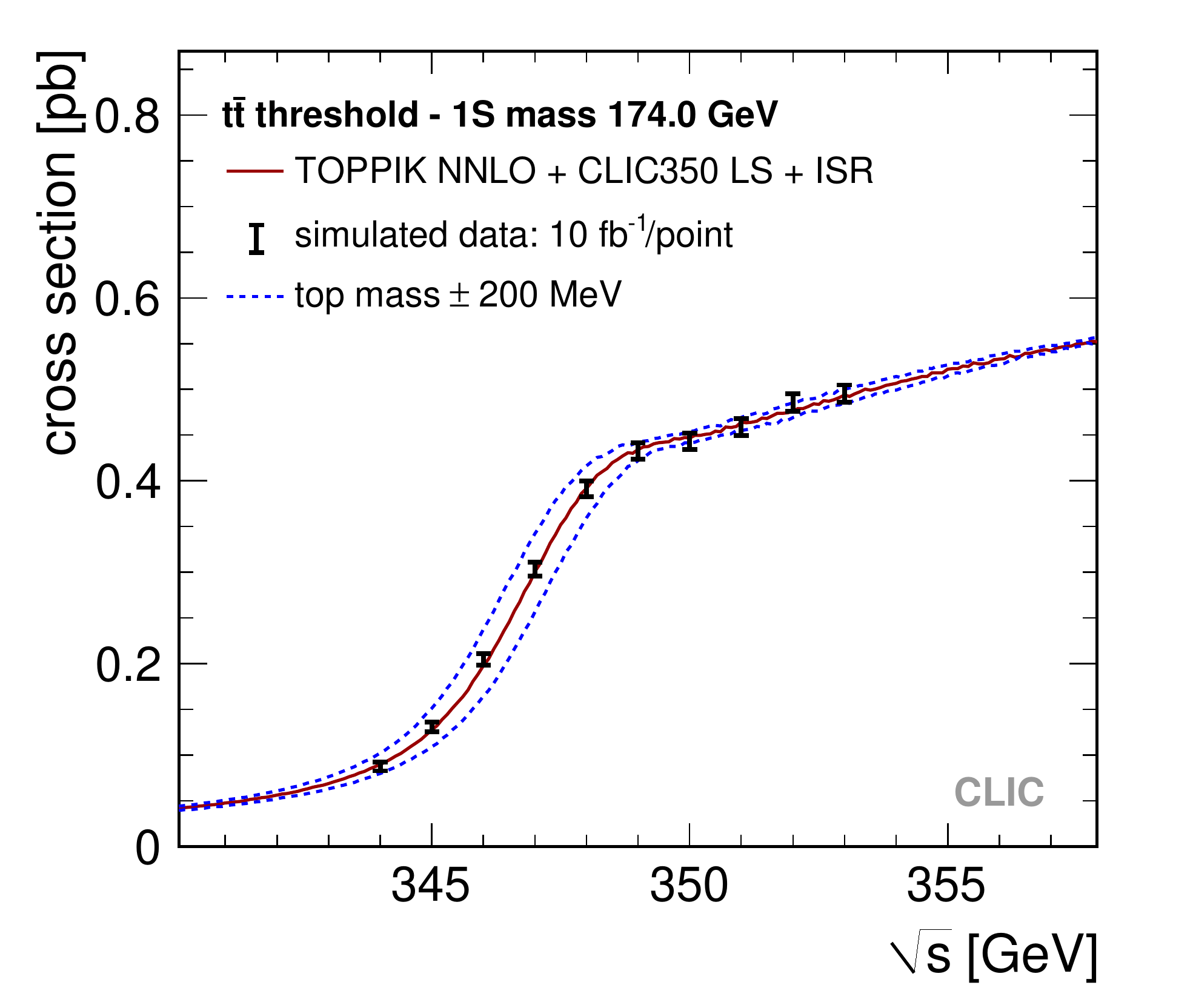}
	\caption{\Leftfig: Reconstructed top quark mass in the all-hadronic decay channel for an integrated luminosity of 100~\fbinv at 500 \GeV. The top mass and width is determined with an unbinned maximum likelihood fit to the invariant mass distribution, shown by the solid line. \Rightfig: Illustration of a scan of the top quark pair production threshold, with each point corresponding to 10~\fbinv of integrated luminosity. The sensitivity to the top quark mass is illustrated by showing the cross section also for 200 MeV changes in mass.}
   \label{fig:Top:Mass}
 \end{figure}

The prospects for top mass measurements at CLIC using both techniques have been studied in full simulations, taking beam-induced and non-\ttbar physics background as well as realistic luminosity spectra into account \cite{Seidel:2013sqa}. With \ttbar cross sections of 450~fb at 350~\GeV and 530~fb at 500~\GeV, integrated luminosities of 100~\fbinv result in several ten thousand \ttbar pairs, enabling  measurements with high statistical precision. 
For both techniques, the two dominating decay modes of the \ttbar pair are considered, the fully-hadronic decay $\ttbar \to \PWp \PQb \PWm \PAQb \to \PQq\PAQq\PQb\,  \PQq\PAQq\PAQb$ and the semi-leptonic decay $\ttbar \to \PWp \PQb \PWm \PAQb \to \PQq\PAQq\PQb\,  \Pl\PGn\PAQb$, where \PGt final states are rejected. Together, these decays account for a branching fraction of 75\% and provide events that can be precisely identified and reconstructed. Top pair identification and reconstruction is performed by a combination of flavor tagging, kinematic fitting and multivariate background rejection, resulting in highly pure signal samples. \autoref{fig:Top:Mass} (\leftfig) shows the reconstructed invariant mass of fully-hadronic top pair decays, together with the remaining non-\ttbar background for an integrated luminosity of 100 \fbinv at an energy of 500 \GeV. The mass is determined with a maximum likelihood fit also shown in the figure. \autoref{fig:Top:Mass} (\rightfig) illustrates a ten-point scan of the \ttbar production threshold corresponding to a total integrated luminosity of 100 \fbinv. The sensitivity to the 1S top mass is illustrated by showing the variations of the threshold behavior for different masses. The top quark mass and the strong coupling constant are extracted simultaneously from a two-dimensional template fit of the measured cross section. \autoref{tab:top} summarizes the achieved statistical precision. In addition to these statistical uncertainties, possible systematic uncertainties have been evaluated. For the invariant mass it was found that jet energy scale systematics can be constrained to a level comparable to the statistical uncertainty. However, since the measurement itself is performed in the context of a leading order event generator, the interpretation of the result in theoretically well-defined mass schemes results in substantial additional uncertainties, expected to be larger than the purely experimental uncertainties quoted here. For the threshold scan, a statistical uncertainty of the top quark mass in the 1S scheme of 33~\MeV is obtained, resulting in a total uncertainty of approximately 100~\MeV when including theoretical normalisation uncertainties as well as analysis-related and beam energy systematics. Additional theory uncertainties on the order of 100~\MeV enter when transforming the 1S mass used in the threshold scan analysis to the $\overline{\mathrm{MS}}$ mass scheme commonly used in electroweak precision calculations. Given the dominance of systematic uncertainties in the mass determination via a threshold scan,  the difference between different \epem collider options is not expected to impact visibly on the total uncertainty \cite{Seidel:2013sqa}.

\begin{table}[tb]
\centering
\caption{Summary of full detector simulation results obtained under
 realistic CLIC beam conditions in the top quark studies. }
\label{tab:top}
\begin{tabular}{l l l l l l l}
  \toprule
  $\roots$ & \multirow{2}{*}{Technique} &Measured & Integrated             & \multirow{2}{*}{Unit} & Generator & Stat. \\
  (\GeV)    &                              &             quantity       &  luminosity ($\fbinv$) &                       & value     & error   \\\midrule
  
  \multirow{2}{*}{350} &  \multirow{2}{*}{Threshold scan} &  Mass & \multirow{2}{*}{$10\times 10$} & \GeV & 174 & 0.033 \\
                       &                                  & $\alpha_\mathrm{S}$ &                          &     & 0.118 & 0.0009\\\midrule

  500                  &  Invariant mass                  & Mass & 100 & \GeV & 174 & 0.080 \\

  \bottomrule
\end{tabular}
\end{table}

\subsection{Top as a Probe for New Physics}

The high mass of the top quark and its correspondingly strong coupling to the mechanism of electroweak symmetry breaking make it a promising probe for New Physics.  For example, the measurement of forward--backward and left--right asymmetries, where the latter makes use of polarized beams provided by a linear collider, provides a high sensitivity to extra-dimensional models and new heavy gauge bosons by probing the \ensuremath{\PQt\PAQt\PZ} and  \ensuremath{\PQt\PAQt\PGg} vertices with high precision. The fact that the top quark decays before it hadronizes, allowing access to its polarization by the analysis of angular distributions of the decay products, makes it a sensitive probe for the couplings to gauge bosons and provides the possibility to search for CP violation in the top sector.

So far, no full simulation studies on these topics have been carried out in the context of CLIC, but the sensitivity of asymmetry measurements to different New Physics models has been studied extensively for the ILC \cite{Doublet:2012wf}. These studies achieve a 1\% precision on the couplings, providing sensitivity for example to Kaluza--Klein particles up to masses of tens of TeV\@. The higher energy at CLIC will allow one to eliminate uncertainties in the assignment of \PQb jets to \PW bosons due to the higher boost of the quarks and the corresponding better separation of the decay products of the \PQt and \PAQt, one of dominating experimental challenges at the ILC at 500 \GeV.  This may result in an increased physics reach, but still requires dedicated studies. Another benefit of the high center-of-mass energy of CLIC is that in most approaches to New Physics affecting the top quark observables, the relative contribution from New Physics increases with the interaction energy $E$ as $E^2/\Lambda^2$, where $\Lambda$ represents the scale of New Physics. Examples for this are the Kaluza--Klein excitation scale, the mass of an extra gauge boson, or the suppression scale of a higher-dimensional operator contributing to \ttbar production.

The high granularity of the detectors together with the particle flow event reconstruction enables the clean reconstruction also of highly boosted top quarks, with a measurement of the individual decay products from an analysis of the jet substructure. The availability of these techniques forms the basis for precision physics with top quarks in the multi-TeV regime. Topics to be investigated in detail in the future include the study of top production asymmetries, the couplings to \PGg, \PZ, \PW and \PH bosons, the sensitivity to CP violation in the top sector and flavor changing top decays. 

\subsection{Conclusion}

Top physics is an integral part of the CLIC physics program. The first stage will provide a precise measurement of the top quark mass on the 100 \MeV level and measurements of other top properties such as the width, while higher energy stages give access to various measurements sensitive to New Physics. The achievable precision for the mass measurements has already been investigated in detail in full simulations, including incomplete studies of systematic uncertainties. The potential to use top quarks as a probe for New Physics will be the subject of studies in the near future, which will include:
\begin{itemize}
\item top quark production asymmetries;
\item top couplings to  \PGg, \PZ, \PW and \PH bosons;
\item CP violation in the top sector;
\item flavor changing top decays.
\end{itemize}

\bibliocommand

%%% Local Variables: 
%%% mode: latex
%%% TeX-master: "../clicsnowmass"
%%% TeX-PDF-mode: t 
%%% End: 

\section{BSM Searches}
\label{sec:bsm-searches}

\subsection{Introduction}
\label{sec:bsm-intro}

It is generally acknowledged that the Standard Model is not the complete picture of particle physics. The quest for a deeper understanding of dark matter, baryon asymmetry of the universe, CP violation, the flavor problem, unification, and stability of the Higgs sector gives rise to a wide spectrum of ideas that extend our theories beyond what the Standard Model can provide.

In this section we give a few examples of how a high-energy \epem CLIC machine can discover and study physics beyond the Standard Model. In \autoref{sec:susy} we discuss the discovery potential and study prospects for supersymmetry, which remains a leading idea of physics beyond the Standard Model. In \autoref{sec:composite} we consider the possibility of a composite Higgs boson, and compare the search sensitivities for its composite nature at CLIC versus LHC\@. In \autoref{sec:exotics} we briefly discuss how additional exotic physics cases can be discerned through careful measurements of high-energy final states. Our primary example is that of a \Zprime boson.

\subsection{Supersymmetry}
\label{sec:susy}

Supersymmetry was posited many years ago as a natural extension of the spacetime structure bearing a symmetry between bosons and fermions. Over time it was recognized that there are many other features that speak for its existence, including having a natural dark matter candidate, revealing a possible unification of the forces at high energies, and having the ability to solve the electroweak scale hierarchy problem. This latter consideration has been under some strain given that the LHC has not yet found superpartners, and it has found a Higgs boson mass of 125~GeV, which may appear to be unnaturally heavy for the minimal model. Nevertheless, these two facts are mutually compatible and higher energy LHC runs are needed to cover more ground in the supersymmetry parameter space.

If supersymmetry is found it will be of primary importance to study all the masses and couplings to high precision to test the many ideas of how supersymmetry can be composed. CLIC offers this opportunity. \autoref{tab:susy_summary_table} shows the excellent precision by which one can measure the superpartner mass spectrum at CLIC\@. The table includes results from the example models \textit{I}, \textit{II}, and \textit{III} which are detailed in the CDR~\cite{CLIC_PhysDet_CDR,CLICCDR_vol3,lcd:2011-016,lcd-model-3}. Accessible states are measured to better than a few percent uncertainty with standard assumptions on the energy and luminosity of CLIC\@.  See the table caption for more details. 

\begin{figure}[tb]
  \centering
  \includegraphics[width=\halfwidth, trim=0cm 0cm 0cm 10cm]{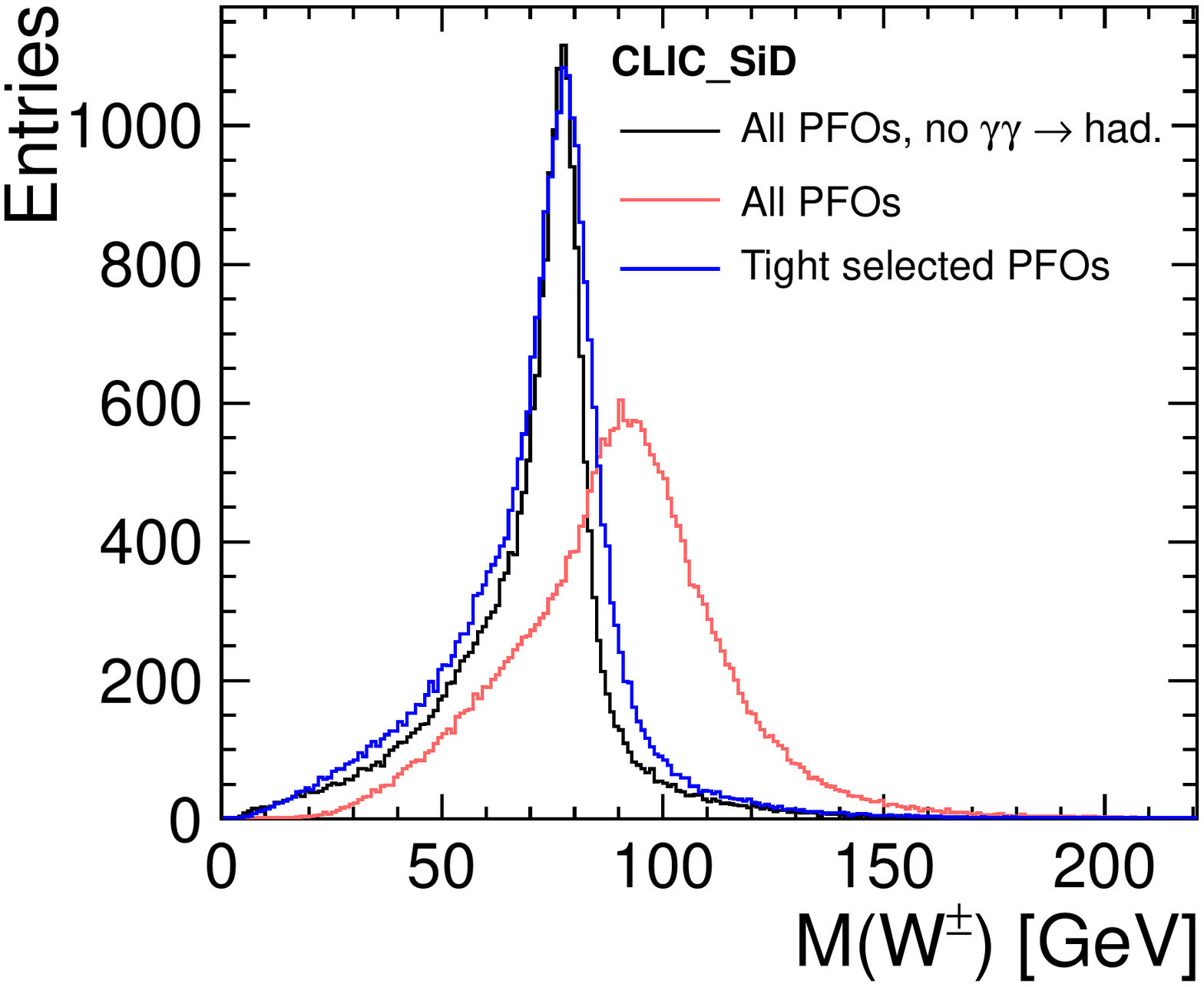}
  \includegraphics[width=\halfwidth, trim=0cm 0cm 0cm 10cm]{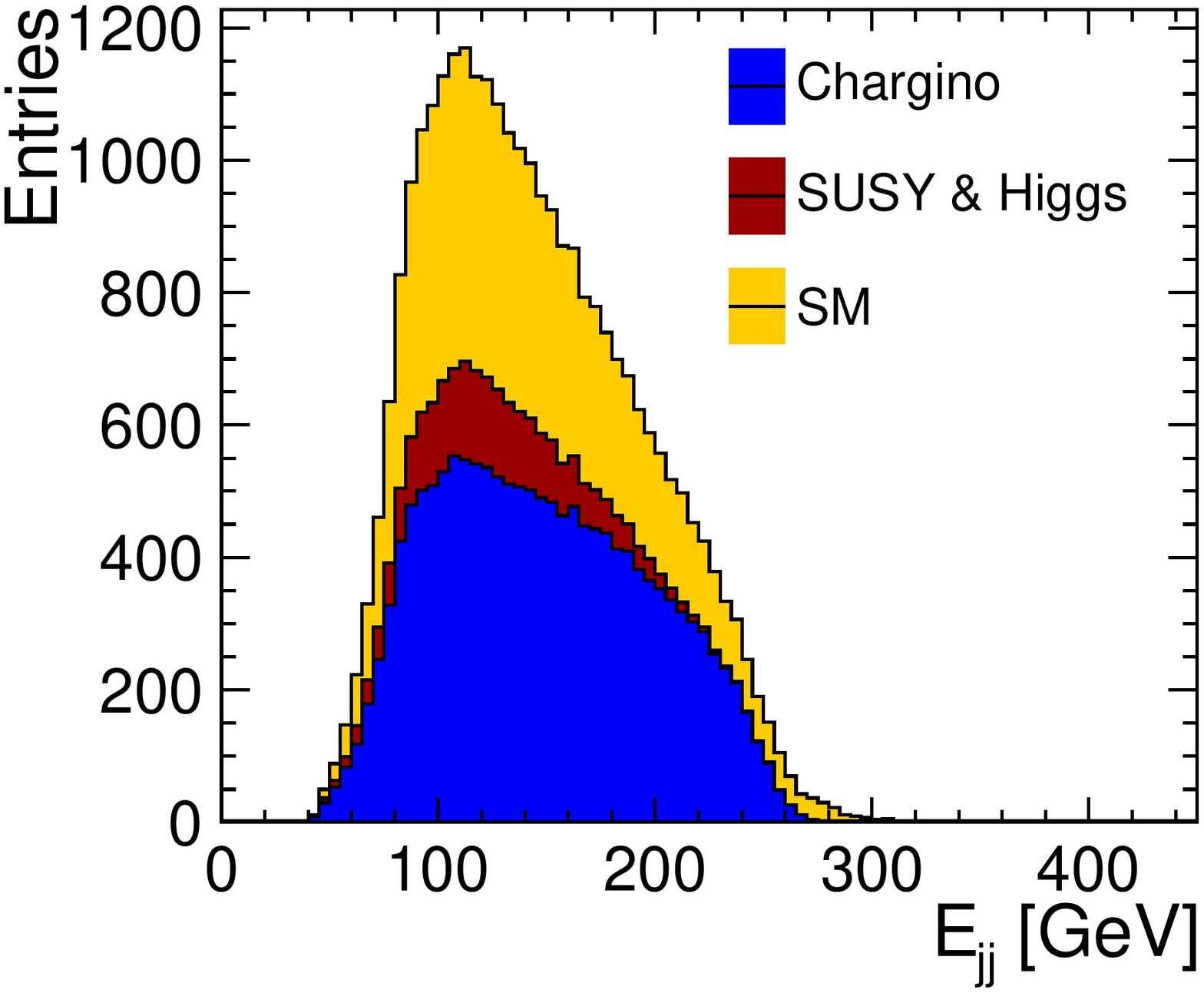}
\caption{\Leftfig: Reconstructed mass of \PWpm candidates in $\epem \to \chch{1}{1} \to \qqqqWW$ events at 1.4~TeV without overlay of \gghadrons (black histogram), with overlay of \gghadrons (red histogram) and using selected reconstructed particles (blue histogram). \Rightfig: Di-jet invariant mass distribution for the selected signal and background events in the same channel. A full simulation of the CLIC\_SiD detector is used. All distributions are scaled to an integrated luminosity of 1.5~\abinv.\label{fig:bsm_gauginos_1400}}
\end{figure}

CLIC allows one to perform precise measurements of the superpartners even using fully hadronic final states which are very difficult at hadron colliders. \autoref{fig:bsm_gauginos_1400} shows an example for a final state with four jets and missing energy. This figure also demonstrates the suppression of pileup from beam-induced backgrounds as introduced in \autoref{sec:introduction}. The mass and pair production cross section for the lightest chargino were extracted from the reconstructed \PW energy distribution in $\epem\to\PSGcp\PSGcm\to\qqqqWW$ events at 1.4~TeV\@. Statistical precisions of 0.2\% for the mass of 487~GeV and 1.3\% for the pair production cross section of 15.3~fb were obtained assuming an integrated luminosity of 1.5~\abinv.

% \autoref{fig:bsm_gauginos_1400} shows an example of how well the superpartner mass spectrum can be measured. The mass and pair production cross section for the lightest chargino were extracted from the reconstructed \PW energy distribution in e$^{+}$e$^{-} \rightarrow \tilde{\chi}^{+}_{1}\tilde{\chi}^{-}_{1} \rightarrow$ q$\bar{\text{q}}$q$\bar{\text{q}}$W$^{+}$W$^{-}$ events at 1.4~TeV. This study demonstrates that hadronic decays of \PW bosons can be well reconstructed despite the pileup from beam-induced backgrounds at CLIC. Statistical precisions of 0.2\% for the mass and 1.3\% for the pair production cross section were obtained assuming an integrated luminosity of 1.5~\abinv.

\begin{figure}[tbp]
  \centering
  \includegraphics[width=\halfwidth]{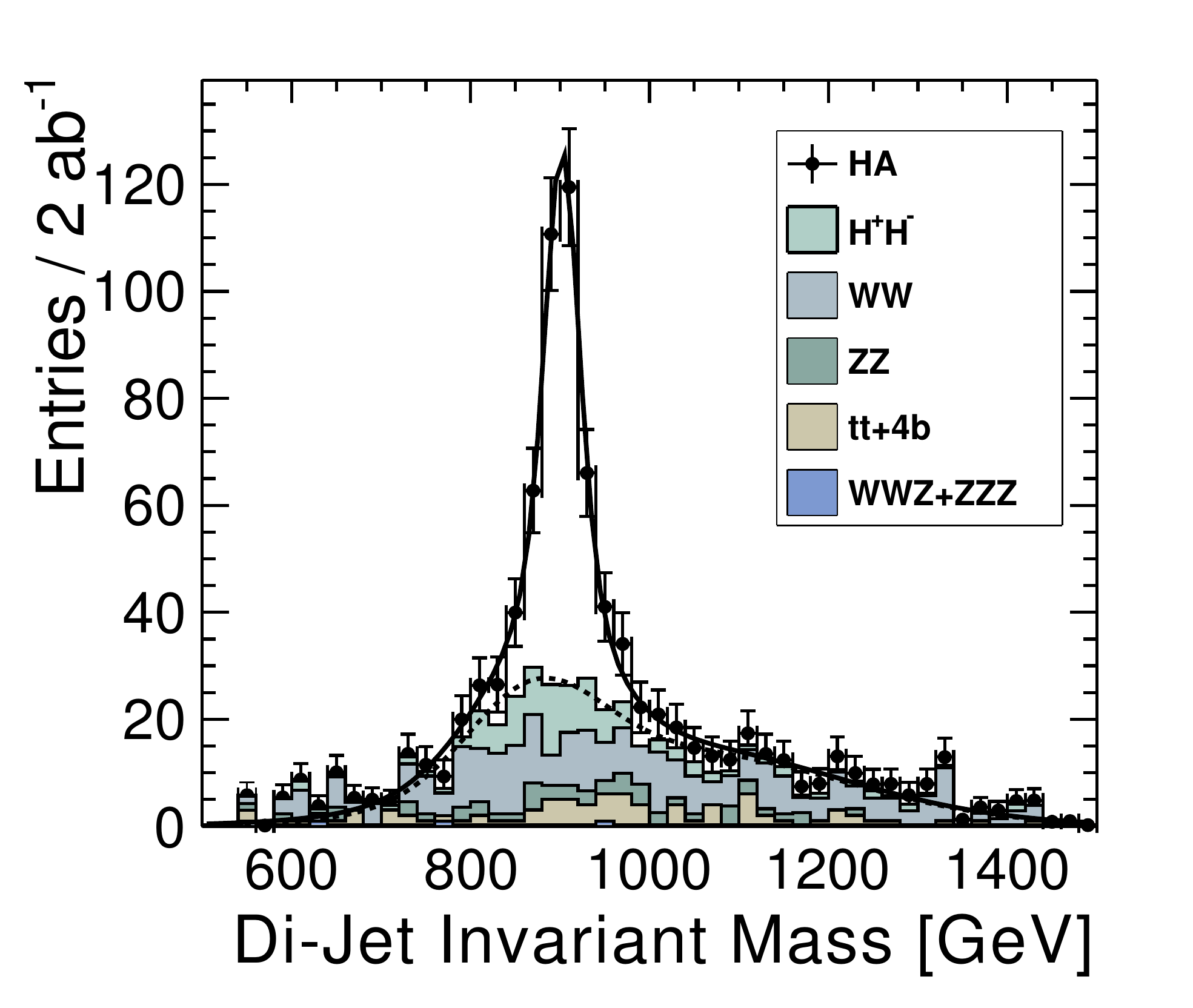}
  \includegraphics[width=\halfwidth]{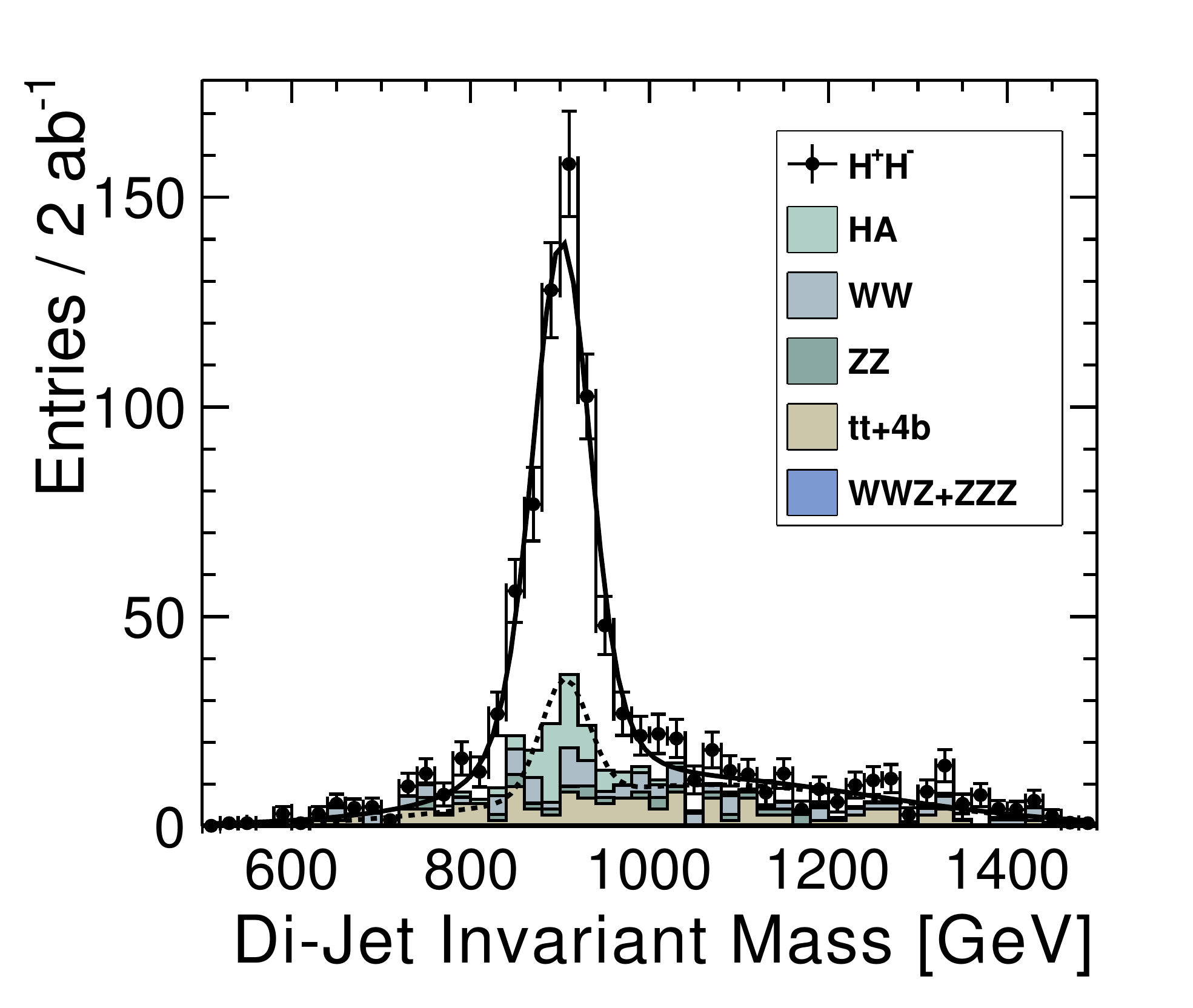}
  \caption{Di-jet invariant mass distributions for the \bbbb (left) and \tbtb (right) final states at 3~TeV for \modelone. The distributions for the $\epem \to \PSH\PSA$ and $\epem\to \PSHp\PSHm$ processes and for the individual backgrounds are shown separately. A full simulation of the CLIC\_ILD detector is used. All distributions are scaled to an integrated luminosity of 2~\abinv.\label{fig:bsm_heavy_higgs}}
\end{figure}

Supersymmetry is necessarily (at least) a Two-Higgs-Doublet theory, and a full demonstration of the theory and a full test of its underlying structure requires measuring the four heavier Higgs degrees of freedom, \PSHpm, \PSA and \PSH. \autoref{fig:bsm_heavy_higgs} shows the ability to measure these masses to the percent level, and to distinguish the mass splitting among all of these states, which can be crucial for understanding the underlying model. 

It should be noted also that two Higgs doublet theories are not necessarily restricted to supersymmetry, therefore this study has a much broader implication than just supersymmetric Higgs boson searches. Indeed, in a broader sense all of these searches have wider applicability than only supersymmetry, since the particles under consideration can be classified simply as states with given mass, spin and quantum numbers, which can show up in any theory, supersymmetric or not.

\subsection{Composite Higgs Boson Theories}
\label{sec:composite}

Since a fundamental scalar boson has quadratic sensitivities to higher scales it is susceptible to quadratically divergent corrections that can destabilize its potential and small mass. Supersymmetry solves this problem via a symmetry, but there are other ways to attempt to solve the problem. A favored option that is not excluded by current LHC data is that the Higgs boson is not a fundamental scalar, but rather a composite state of fermions. In this way electroweak symmetry breaking and mass generation are at their essence a condensation of multiple fields together, whose composite nature mimics a condensing scalar boson -- the Higgs boson.

The phenomenology of composite Higgs theories is very similar to the phenomenology of the SM Higgs boson. The only difference is that every observable has relative corrections to it that are proportional to $\xi=(v/f)^{2}$, where $v\simeq 246~\GeV$ is the normal vacuum expectation value of the ``Higgs'', and $4\pi f$ is the (higher) scale of compositeness. The scale $f$ cannot be too small otherwise corrections to normal Higgs production and decay are $\xi={\cal O}(1)$ which is forbidden by current data. Therefore, $\xi\ll 1$ is required. Exactly how low needs to be derived carefully from data.

\begin{figure}[t]
  \centering
  \includegraphics[width=0.45\linewidth]{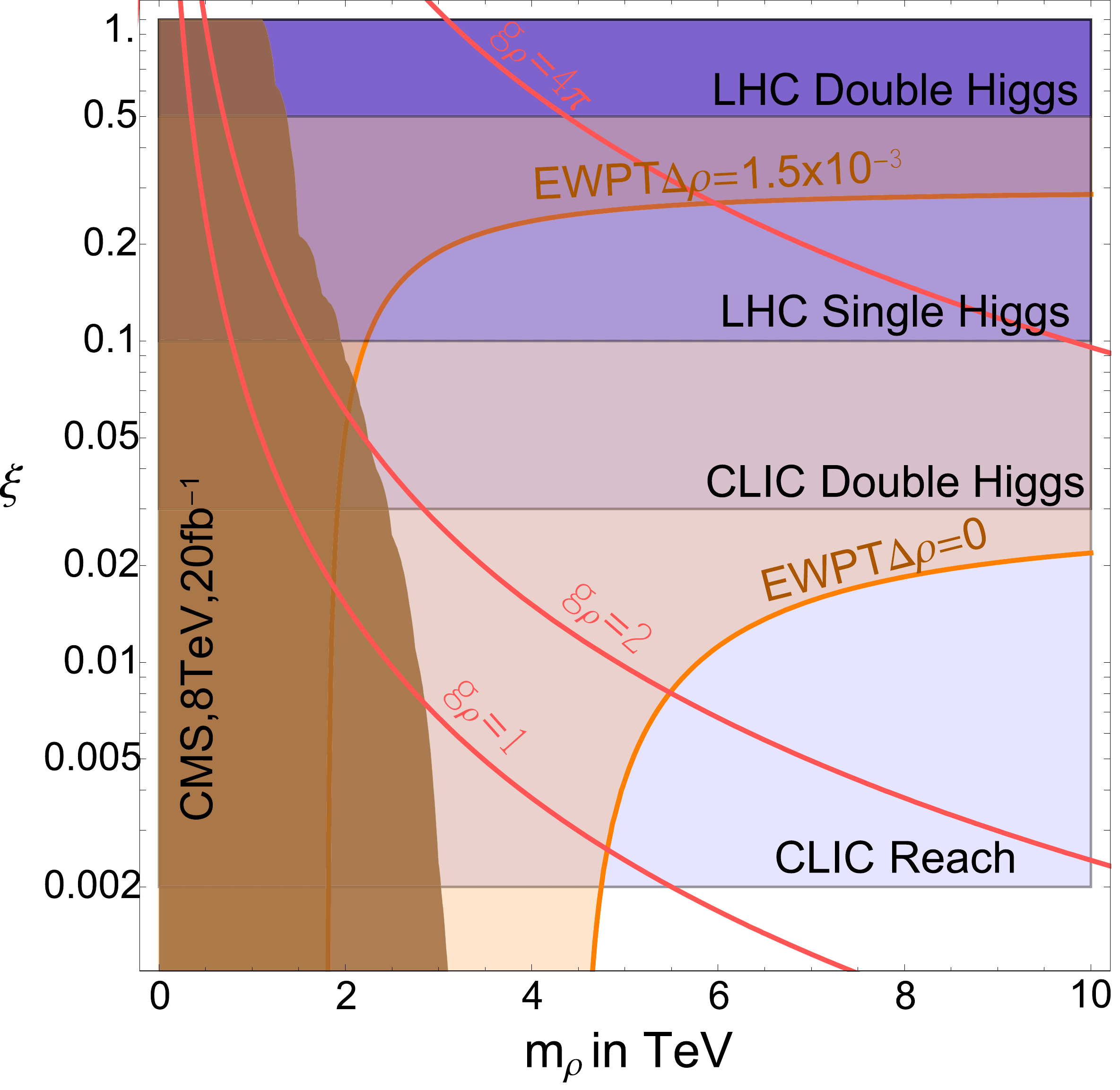}
\caption{Summary plot of the current constraints and prospects for direct and indirect probes of the strong interactions triggering electroweak symmetry breaking. $m_{\rho}$ is the mass of the vector resonances and $\xi = (v/f)^{2}$ measures the strengths of the Higgs interactions. The dark brown region on the left shows the current combined limit from direct production of the charged $\rho^\pm$ at the LHC decaying to $\ell\upnu$ and $\PW\PZ \to 3\ell\upnu$ final states. The dark (medium light) horizontal purple bands indicate the sensitivity on $\xi$ expected at the LHC from double (single) Higgs production with 300~\fbinv of integrated luminosity. 
The pink horizontal band reports the sensitivity reach on $\xi$ from the study of double Higgs processes alone at CLIC with 1~\abinv of integrated luminosity at 3~TeV while the light blue horizontal band shows the sensitivity reach on $\xi$ when considering single Higgs processes.
%The pink horizontal band reports the sensitivity reach on $\xi$ from the study of double Higgs processes alone at CLIC with 1~\abinv of integrated luminosity at 3~TeV. 
Finally, experimental electroweak precision tests (EWPT) favor the region below the orange thick line with and without additional contribution to $\epsilon_1$. The Higgs mass is assumed to be 125~GeV and the vector resonance contribution to $\epsilon_3$ is taken to be $\Delta \epsilon_3 = m_{\PW}^{2}/m_{\rho}^{2}$. The domain of validity of our predictions, $g_\rho < 4\pi$, is below the upper red line (From~\cite{contino13:strongHiggslc}).
\label{fig:bsm_composite_higgs}}
\end{figure}

\autoref{fig:bsm_composite_higgs} (from ~\cite{contino13:strongHiggslc}) shows the sensitivities at LHC and CLIC for observing non-SM signatures from the composite nature of the Higgs boson in the plane of $\xi$ and $m_\rho$, where $\rho$ is the vector resonance of the composite theory, in direct analogy to the $\rho$ of QCD which regularizes composite pion phenomenology. A detailed description of \autoref{fig:bsm_composite_higgs} is given in the caption. With an integrated luminosity of 1~\abinv accumulated at 3~TeV, CLIC can reach $\xi \approx 0.03$ independent of $m_{\rho}$ due to the relatively clean environment for studying double Higgs boson production. Note that the reach on $\xi$ is improved by about one order of magnitude, down to $\xi \approx 0.002$, thanks to the precise measurement of Higgs couplings in single Higgs processes, as reported in \autoref{tab:higgs:kappas} of this document. This will be an indirect but powerful probe of a Higgs composite scale up to 70 TeV.
% The reach on $\xi$ can be improved significantly when combining precision results from the combined CLIC Higgs coupling fit with double Higgs production and was estimated to be around $\xi=0.002$~\cite{contino13:strongHiggslc}, corresponding to a Higgs composite scale of 70~TeV. 
For comparison, the LHC with an integrated luminosity of 300~\fbinv at 14~\TeV can reach only down to $\xi \approx 0.1$, given the most recent estimates. 

\subsection{Search for Exotic Physics through Direct Production and Precision Studies}
\label{sec:exotics}

\begin{figure}[t]
\centering
\includegraphics[width=0.75\textwidth,clip,trim=40 30 45 30]{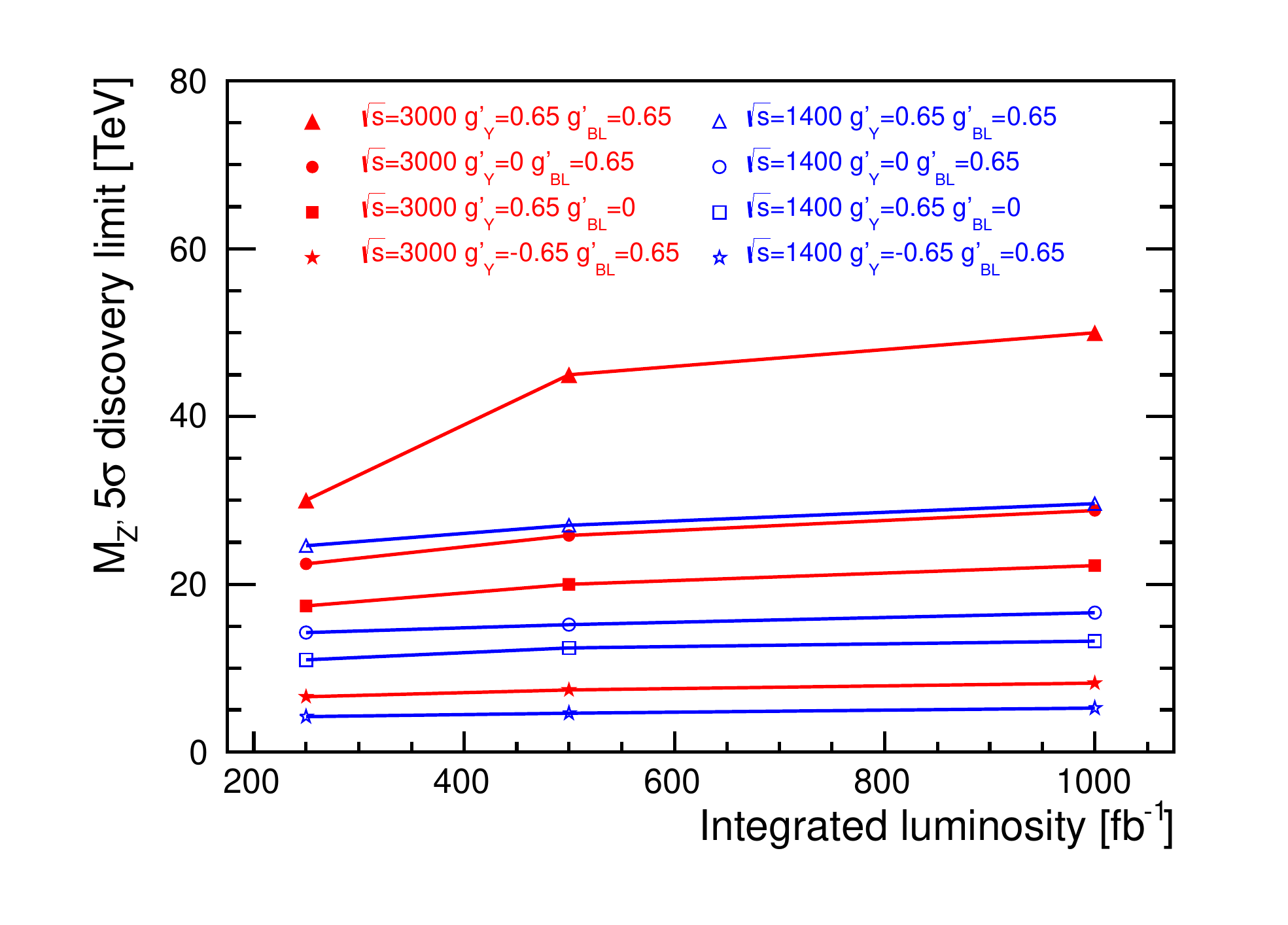} \hfill
\caption{$5\sigma$ limit for a $M_{\Zprime}$ discovery as function of the integrated luminosity for different values of the couplings $g'_{\mathrm{Y}}$ and $g'_{\mathrm{BL}}$. The limits shown are determined from the combined observables $\sigma$ and $A_{\mathrm{FB}}$ at $\sqrt{s} = 3~\TeV$ and 1.4~\TeV.\label{fig:bsm_zprime}}
\end{figure}

CLIC is a precision \epem machine, and as such it is able to study many different observables with sub-percent accuracy. For example $\epem\to\fbarf$ observables can be key to seeing small deviations with respect to the SM when lepton scattering energies are in the TeV region.  The observables include total cross section, forward--backward asymmetry, and polarization asymmetries. In general one can put strong constraints on higher-dimensional operators that connect electrons to muons, for example, from these precision studies of $\epem \to \mpmm$. A concrete case of this impressive general capacity is sensitivity to a new \Zprime that couples to leptons~\cite{Blaising:2012tz}. The new $\Zprime$ is the most general $U(1)'$ gauge symmetry that is anomaly-free with respect to the SM particle content. Under this $U(1)'$ the SM fermions have charge $g'_\mathrm{Y}(Y_\mathrm{f})+g'_{\mathrm{BL}}(B-L)_\mathrm{f}$.  \autoref{fig:bsm_zprime} shows $5\sigma$ discovery limits for \Zprime gauge boson mass as a function of the achieved integrated luminosity of the machine using the measured cross section and asymmetries.
It was found, for example, that there are regions of parameter space for this anomaly free \Zprime theory where sensitivity of the mass reaches into several tens of TeV\@, well beyond the center-of-mass energy of the machine, and well beyond what the LHC or its conceived upgrades can achieve. This impressive result is characteristic of many such studies that can be mapped to a non-renormalizable operator that connects the electrons to any ``clean'' final state, such as muons or gauge bosons.

\subsection{Conclusion}

In our discussion of some of the leading ideas of physics beyond the SM we have found significant discovery potential. In addition, \autoref{tab:bsm_discovery_reach} gives a brief survey of discovery reach of the CLIC collider compared to the LHC and its possible high-luminosity upgrades in various other beyond the SM theories. It is adapted from the CLIC CDR~\cite{CLIC_PhysDet_CDR}. As illustrated in our examples earlier in the text and in \autoref{tab:bsm_discovery_reach}, some indirect precision studies allow the discovery of signals beyond the SM originating from scales well beyond the center-of-mass energy of the collider.  We have also seen that if particles are discovered at CLIC or the LHC, CLIC has the ability to measure the masses and couplings at the percent level or better. This is generally far more precise than what a hadron collider can do alone. The precision studies complementary to the LHC and the stand-alone discovery and precision capacity of CLIC makes it an ideal machine for extending our search for physics beyond the SM\@. Future effort in beyond the SM physics will include
\begin{itemize}
\item Searches for dark matter missing energy signatures in a model-independent way;
\item Searches and study of resonances associated with composite Higgs theory;
\item Generalization of higher-dimensional effective operator searches at the various stages of CLIC running;
\item Searches for very weakly interacting exotic particles;
\item Searches for vectorlike particles charged under electroweak group;
\item Responding to theory guidance for New Physics that is compatible and explains LHC data in the future.
\end{itemize}

% In our brief survey of some of the leading ideas of physics beyond the SM we have found significant discovery reach capacity. \autoref{tab:bsm_discovery_reach} gives a brief survey of discovery reach capabilities of the CLIC collider compared to other colliders that are presently operating or conceived. It is adapted from the CLIC CDR~\cite{CLIC_PhysDet_CDR}. In some indirect precision studies the discovery of signals beyond the SM can originate from physics well beyond the center-of-mass energy of the collider. We have also seen that if particles are discovered at the CLIC or the LHC, CLIC has the ability to measure the masses and mixing angles at the percent level of better, which generally is far more precise than what a hadron collider alone can do. The precision studies complementary to the LHC and the stand-alone discovery and precision capacity of CLIC makes it an ideal machine for extending or search for beyond the SM physics.

\begin{table}[htbp]
\vspace{2cm}
\centering
\caption{Summary table of the CLIC SUSY benchmark analyses results obtained with full-detector simulations with background overlaid. All studies are performed at a center-of-mass energy of 3 TeV (1.4~TeV) and for an integrated luminosity of 2~\abinv (1.5~\abinv)~\cite{Battaglia:2010in, lcd:2011-018, LCD-2011-027, LCD:2011-37, LCD:2012-004, LCD:2012-006, lcd:2012-012}.\label{tab:susy_summary_table}}
\begin{tabular}{l l l l l l l}
\toprule
$\sqrt{s}$ & Process & Decay mode & SUSY & Measured & Generator & Stat. \\
(TeV)    &         &            & model &      quantity   & value (GeV) &uncertainty \\
\midrule
\midrule

	%------------------sleptons production-------------------------------------
        \multirow{6}{*}{3.0} & \multirow{6}{*}{Sleptons} &  \multirow{2}{*}{$\PSGmpR\PSGmmR\to\mpmm\,\nene{1}{1}$} & \multirow{6}{*}{II} & $\tilde\ell$ mass &1010.8 &  0.6\% \\
        &            & & &  $\neutralino{1}$ mass & 340.3 & 1.9\%\\

	%-------------------------------------------
	& & \multirow{2}{*}{$\PSepR\PSemR\to \epem\,\nene{1}{1}$}  & & $\tilde\ell$ mass & 1010.8 & 0.3\% \\
        & & & & $\neutralino{1}$ mass & 340.3 & 1.0\%\\
 	
	%-------------------------------------------
	& & \multirow{2}{*}{$\PSGne\PSGne\to \nene{1}{1}\epem\ww$}  & & $\tilde\ell$ mass & 1097.2 & 0.4\% \\
        & & & & $\chargino{\pm}$ mass& 643.2 & 0.6\%\\
 
	%--------------chargino and neutralino production-------------------------------
	\midrule
	\multirow{2}{*}{3.0} & Chargino & $\chch{1}{1}\to\nene{1}{1}\ww$ & \multirow{2}{*}{II} & $\chargino{\pm}$ mass& 643.2 & 1.1\% \\
	& Neutralino & $\nene{2}{2} \to \Ph/\PZz\;\Ph/\PZz \; \nene{1}{1}$ &  & $\neutralino{2}$ mass & 643.1 & 1.5\% \\

	%-----------------production of right-handed squarks --------------------------
	\midrule
	3.0 & Squarks & $\PSQR\PSQR \to \PQq\PAQq \nene{1}{1} $ & I & $\PSQR$ mass & 1123.7 & 0.52\% \\

	%----------------heavy Higgs production ---------------------------------------
	\midrule
	\multirow{2}{*}{3.0} & \multirow{2}{*}{Heavy Higgs} & $\PSHz\PSAz \to \bbbb$ & \multirow{2}{*}{I} & $\PSHz/\PSAz$ mass  & 902.4$/$902.6 & 0.3\% \\
	                     &             & $\PSHp\PSHm \to \tbtb$ &                    & $\PSHpm$ mass  & 906.3 & 0.3\% \\
	
        \midrule

        \multirow{6}{*}{1.4} & \multirow{6}{*}{Sleptons} &  \multirow{2}{*}{$\PSGmpR\PSGmmR\to \mpmm\,\nene{1}{1}$} & \multirow{6}{*}{III}  & $\slepton{}$ mass     & 560.8 &  0.1\% \\
                             &            &                                                                        &                       &  $\neutralino{1}$ mass & 357.8 & 0.1\%  \\

	%-------------------------------------------
	                     &            & \multirow{2}{*}{$\PSepR\PSemR\to \epem\,\nene{1}{1}$}  &                       & $\tilde\ell$ mass      & 558.1 & 0.1\%  \\
                             &            &                                                                        &                       & $\neutralino{1}$ mass  & 357.1 & 0.1\%  \\
 	
	%-------------------------------------------
	                    &             & \multirow{2}{*}{$\PSGne\PSGne\to \nene{1}{1}\epem\ww$}&                       & $\tilde\ell$ mass      & 644.3 & 2.5\% \\
                            &             &                                                                        &                       & $\chargino{\pm}$ mass  & 487.6 & 2.7\% \\
\midrule

	\multirow{1}{*}{1.4} & Stau       & $\PSGtpDo\PSGtmDo\rightarrow \PGtp\PGtm\nene{1}{1}$ & III & $\PSGtDo$ mass & 517 & 2.0\% \\

\midrule

	\multirow{2}{*}{1.4} & Chargino   & $\chch{1}{1}\to\nene{1}{1}\ww$ & \multirow{2}{*}{III} & $\chargino{\pm}$ mass& 487  & 0.2\% \\
	                     & Neutralino & $\nene{2}{2} \to \Ph/\PZz\;\Ph/\PZz \; \nene{1}{1}$ &                      & $\neutralino{2}$ mass& 487  & 0.1\% \\

\midrule
\end{tabular}
\end{table}

\begin{table}[htbp]
\centering
\caption{Discovery reach of various theory models for different colliders~\cite{CLIC_PhysDet_CDR}. LHC at $\sqrt{s}$ = 14~TeV assumes 100~\fbinv of integrated luminosity, while HL-LHC is with 1~\abinv, and CLIC3 is $\sqrt{s} = 3~\TeV$ with up to 2~\abinv. TGC is short for Triple Gauge Coupling, and ``$\mu$ contact scale'' is short for LL $\mu$ contact interaction scale $\Lambda$ with $g=1$.\label{tab:bsm_discovery_reach}}
\begin{tabular}{cccc}
\toprule
New particle &     LHC (14~TeV) & HL-LHC & CLIC3\\
\midrule
\midrule
squarks [TeV] &   2.5 & 3 & $\lesssim$1.5 \\
sleptons [TeV] &   0.3 & - & $\lesssim$1.5 \\ 
\Zprime ({\tiny SM~couplings}) [TeV]  &  5 & 7 & 20   \\ 
2 extra dims $M_D$ [TeV]  &    9 & 12 & 20--30 \\
TGC (95\%)  ({\tiny \rm $\lambda_{\gamma} $~coupling}) &   0.001 & 0.0006 & 0.0001 \\
$\mu$ contact scale [TeV] &  15& - & 60 \\
Higgs composite scale [TeV] & 5--7 & 9--12 & 70\\
\bottomrule
\end{tabular}
\end{table}

\bibliocommand

%%% Local Variables: 
%%% mode: latex
%%% TeX-master: "../clicsnowmass"
%%% TeX-PDF-mode: t 
%%% End: 

\section{Precision Study of Electroweak Interactions}
\label{sec:prec-meas}

One of the tasks of an \epem collider is to test the Standard Model electroweak predictions with high precision. In \autoref{sec:higgs-physics} we have detailed the expectations of precision measurements for the Higgs boson sector, which is the most pressing electroweak precision program of the linear collider.

Likewise, we have seen that the precision measurement capabilities of CLIC in the presence of New Physics are well supported by recent analyses briefly described earlier in \autoref{sec:bsm-searches}. In the example case of supersymmetry, we see that superpartner masses can be measured to sub-percent accuracy. The masses of exotic heavy Higgs bosons also can be measured to this precision, as demonstrated in \autoref{sec:bsm-searches}. This excellent precision will reflect itself in whatever New Physics may be found in the high-energy frontier.

Another aspect of the precision electroweak program is to measure Standard Model processes to unprecedented levels to test self-consistency of the Standard Model framework. The observables most often considered in this context are the mass of the $\PW$ boson, the weak mixing angle $\sin^2\theta_\mathrm{W}$ measured at various energies for various final state particles, and anomalous triple and quartic gauge boson couplings.  There have been few CLIC studies up to this point on these pure electroweak precision processes in the context of the CLIC collider. However, as is the case with other measurements, we expect the results for lower-energy stages of CLIC to be comparable to the results found for the ILC, and we expect that some improvement can arise at the higher-energy and higher-luminosity phases of CLIC.

Large numbers of $\PW$ bosons will be produced in $\Pe\Pe \to \Pe\PW\PGn$ events at a high-energy CLIC collider. Including the effects from ISR and Beamstrahlung, $22 \times 10^{6}$ events for 1.5~\abinv of data collected at 1.4~TeV and $45 \times 10^{6}$ events for 2~\abinv of data collected at 3~TeV are expected assuming unpolarized beams. It was found in a generator-level study that the number of hadronic $\PW$ decays expected is $9 \times 10^{6}$ at 1.4~TeV and $15 \times 10^{6}$ at 3~TeV. Both jets from the $\PW$ decay were requested to be in the central region (i.e., $|\cos(\theta^{\mathrm{jet}})| < 0.94$). These samples provide the potential for a competitive measurement of the $\PW$ boson mass using its hadronic decays. A full simulation study is foreseen to study the impact of systematic effects like uncertainty of the jet energy scale on this measurement.

Regarding the anomalous triple gauge boson vertices, the CP-even couplings are defined in~\cite{Hagiwara:1986vm,Barklow:1996in} to arise from operators in the Lagrangian
\begin{eqnarray}
\Delta{\cal L} & = & ig_1^V(W^\dagger_{\mu\nu} W^\mu V^\nu-W^\dagger_\mu V_\nu W^{\mu\nu})
+ i\kappa_V W^\dagger_\mu W_\nu V^{\mu\nu}+\frac{i\lambda_V}{M_\mathrm{W}^2}W^\dagger_{\lambda\mu}W^\mu_\nu V^{v\lambda} \\
               &   & -g^V_4 W^\dagger_\mu W_\nu(\partial^\mu V^\nu+\partial^\nu V^\mu) +g^V_5 \epsilon^{\mu\nu\rho\sigma}
(W^\dagger_\mu\partial_\rho W_\nu +\partial_\rho W^\dagger_\mu W_\nu)V_\sigma 
\end{eqnarray}
with the normalization $g_1^V=\kappa_V=1$ and $\lambda_V=g_4^V=g_5^V=0$ in the Standard Model at tree level. $\Delta g_1^V\equiv g_1^V-1$ and $\Delta\kappa_V=\kappa_V-1$ define shifts away from the Standard Model values.

The anomalous triple gauge couplings have been studied for CLIC in the past~\cite{Accomando:2004sz}. These studies are not at the same mature full-simulation level that the more recent CLIC studies engage; however, they are expected to be reasonable estimates of what can be achieved in the high-energy CLIC environment. \autoref{tab:precSens} reproduces the results from~\cite{Accomando:2004sz} from precision $\epem\to \PWp\PWm$ analysis, where the various couplings are defined in~\cite{Accomando:2004sz,Hagiwara:1986vm}.  The superscripts $\mathrm{L}$ ($\mathrm{R}$) refer to the values obtained for amplitudes with left (right) handed electrons and right (left) handed positrons.  The definition for $g_1$ for example is from the combinations
 \begin{eqnarray}
 g_1^\mathrm{L}&=& 4\sin^2\theta_\mathrm{W} g^\gamma_1+(2-4\sin^2\theta_\mathrm{W})g_1^Z\xi \\
 g_2^\mathrm{R}&=& 4\sin^2\theta_\mathrm{W} g_1^\gamma-4\sin^2\theta_\mathrm{W} g_1^Z\xi
\end{eqnarray}
 where $\xi=s/(s-m_{\PZ}^2)$. For more details, see~\cite{Accomando:2004sz}.

The improvements in the sensitivity of these couplings can be derived from statistical arguments and analysis of the cross section scaling. One finds that for the couplings that derive from gauge-invariant dimension-six operators with two derivatives, such as the $\lambda_i$ couplings, the improvement in sensitivity to the anomalous couplings scales as $\sqrt{s\LumiInt}$, where $s$ is the center-of-mass energy squared of the collision and $\LumiInt$ is the integrated luminosity.  The strength of this improvement derives from the fact that the higher dimension operator has derivatives that turn to energy factors in the numerator of the correction factor. Numerically, this implies that the value of 0.59 obtained for $\mathrm{Re}(\lambda_L)$ at $\sqrt{s}=500~\GeV$ becomes 0.26 for the $800~\GeV$ stage and 0.04 for the $3000~\GeV$ stage, which is close to the 0.24 and 0.036 actual values found in~\cite{Accomando:2004sz} and shown here in \autoref{tab:precSens}. The significant improvements of the sensitivity to these couplings with increased energy, as can be seen in this study, has been recognized for some time now~\cite{Barklow:1996in}.

There is not necessarily an equivalently simple statistical scaling argument for the $g_i$ and $\Delta\kappa$ couplings, as their definition is to disrupt the gauge-invariant renormalizable couplings of the Standard Model. A calculation of the energy dependence of the cross section is needed for each stage in these cases. The energy scaling for $\Delta\kappa$ turns out to be similar to that of the $\lambda_i$ couplings, and so significant improvement takes place. The anomalous $g_i$ couplings do not have the same scaling; nevertheless, the analysis of~\cite{Accomando:2004sz} as shown in the table shows that modest improvements of their sensitivities occur with increased energy and luminosity.

\begin{table}[t]
  \centering
  \caption{Sensitivity of the real parts of CP-even couplings in units of
    $10^{-3}$, defined and expounded upon in~\cite{Accomando:2004sz}. The integrated
    luminosities for the $500~\GeV$, $800~\GeV$ and $3000~\GeV$ stages are assumed
    here to be $500~\fbinv$, $1~\abinv$ and $3~\abinv$ respectively.}\label{tab:precSens}
  \begin{tabular}{ccccccccc}
    \toprule
    $\sqrt{s}~[\GeV]$ & $\mathrm{Re}(\Delta g_1^L)$ & $\mathrm{Re}(\Delta \kappa_L) $ & $\mathrm{Re}(\lambda_L)$ & $\mathrm{Re}(g_5^L)$ & $\mathrm{Re}(g_1^R)$ & $\mathrm{Re}(\Delta \kappa_R)$ & $\mathrm{Re}(\lambda_R)$ & $\mathrm{Re}(g_5^R)$ \\\midrule
    500             & 2.6                         & 0.85                            & 0.59                     & 2.0                  & 10                   & 2.4                            & 3.6                      & 6.7                  \\
    800             & 1.6                         & 0.35                            & 0.24                     & 1.4                  & 6.2                  & 0.92                           & 1.8                      & 4.8                  \\
    3000            & 0.93                        & 0.051                           & 0.036                    & 0.88                 & 3.1                  & 0.12                           & 0.36                     & 3.2                  \\
    \bottomrule
  \end{tabular}
\end{table}

The precision electroweak measurements program at CLIC will proceed in the future with full simulations of relevant observables. This activity includes
\begin{itemize}
\item Simulations of triple and quartic gauge boson vertex corrections to $\epem\to \ww (\PGn\PGn/\epem)$;
\item Simulations of the forward--backward and left--right asymmetries of fermion production to achieve precision measurements of $\sin^2\theta^{\mathrm{eff}}_\mathrm{f}$ at various energy stages;
\item Simulations of \PW boson mass determination at high energy and high luminosity;
\item Simulations of total $\epem \to \fbarf$ cross sections at high energy with various electron--positron polarizations in search of form-factor suppressions or enhancements.
\end{itemize}

\bibliocommand

%%% Local Variables: 
%%% mode: latex
%%% TeX-master: "../clicsnowmass"
%%% TeX-PDF-mode: t 
%%% End: 

\section{Summary and Conclusions}
\label{sec:summary-conclusions}

The CLIC accelerator is an attractive option for a future high-energy $\epem$ linear collider operating at center-of-mass energies up to 3~\TeV. 
The feasibility of the CLIC accelerator was demonstrated through extensive prototyping, simulations and large-scale tests, 
as described in the conceptual design report~\cite{CLICCDR_vol1}. The physics reach of CLIC was studied in detail and the majority
of the results described in this document are based on full detector simulation and event reconstruction, taking into account the pile-up of 
background from $\gghadrons$. 

This report summarises the physics potential of CLIC operating in three distinct energy stages. The first stage at $\roots\approx350~\GeV$ provides
precise measurements of the properties of the Higgs boson and the top quark. Subsequent high-energy running, here taken to be
at $\roots=1.4~\TeV$  and  $\roots=3.0~\TeV$, provides the potential to accumulate large samples of Higgs boson decays providing
a range of Higgs boson couplings at the ${\cal{O}}(2\%)$ level, going significantly beyond what is achievable at the HL-LHC\@. 
This level of precision may be necessary to distinguish the light Higgs boson of an extended theory from a Standard Model Higgs boson. 
Furthermore, high-energy CLIC operation allows to measure the Higgs trilinear self-coupling parameter $\lambda$ at the 10\% level. 
In addition to probing the electroweak symmetry breaking 
mechanism, the operation of CLIC at $\roots > 1~\TeV$ would provide sensitivity to a wide range of phenomena beyond the Standard Model, 
complementary to that achievable at the HL-LHC\@. For example, CLIC could provide precise measurements of the non-colored TeV-scale particles of SUSY\@. 
In particular, CLIC would enable pair-produced gaugino, slepton and heavy Higgs boson masses to be measured with ${\cal{O}}(1\%)$ 
precision, with sensitivity extending up to the kinematic limit of $m\approx1.5~\TeV$. In addition to studying new particles directly, 
CLIC provides sensitivity to New Physics through precision measurements, where, for example, $\Zprime$ and Higgs 
compositeness models can be probed up to scales of approximately 20~TeV and 30~TeV respectively.

Given its feasibility, staged implementation and its broad physics program beyond and complementary to HL-LHC, there is a strong case for
CLIC being the next energy-frontier accelerator operating above 1~TeV\@. It will provide exciting research opportunities at the forefront of particle physics for several decades.

\bibliocommand

%%% Local Variables: 
%%% mode: latex
%%% TeX-master: "../clicsnowmass"
%%% TeX-PDF-mode: t 
%%% End: 

\bibliographystyle{Main/cliccdr}
\bibliography{Bibliography/cdrbibliography,Bibliography/lcdbib,Bibliography/snowmass}

\end{document}